\documentclass[preprint,tightenlines,preprintnumbers,aps,eqsecnum,nofootinbib,superscriptaddress,showpacs,floatfix]{revtex4-1}

\usepackage{amsmath}    
\usepackage{bm}         
\usepackage{graphicx}   
\usepackage{verbatim}   
\usepackage{color}      
\usepackage{subfigure}  
\usepackage[colorlinks=true,citecolor=blue,linkcolor=blue,urlcolor=blue]{hyperref} 
\usepackage{mathrsfs}

\begin{document}


\newcommand{\rep}[1]{{\bf #1}}
\newcommand{\be}{\begin{equation}}
\newcommand{\ee}{\end{equation}}
\newcommand{\ben}{\begin{eqnarray}}
\newcommand{\een}{\end{eqnarray}}
\newcommand{\bel}[1]{\be\label{#1}}
\newcommand{\bmini}{\begin{minipage}}
\newcommand{\emini}{\end{minipage}}
\newcommand{\txw}{\textwidth}
\newcommand{\pic}{\includegraphics}
\newcommand{\toright}{\raggedleft}
\newcommand{\toleft}{\raggedright}
\newcommand{\bitem}{\begin{itemize}}
\newcommand{\eitem}{\end{itemize}}

\newcommand{\rhoV}[1]{\ensuremath{\rho_{V^{#1}}}}
\newcommand{\ZV}{\ensuremath{Z_{V^4}}}
\newcommand{\ZVQQ}{\ensuremath{Z_{V^4_{QQ}}}}
\newcommand{\ZVbb}{\ensuremath{Z_{V^4_{bb}}}}
\newcommand{\ZVcc}{\ensuremath{Z_{V^4_{cc}}}}
\newcommand{\ZVbc}{\ensuremath{Z_{V^4_{bc}}}}
\newcommand{\ZVcb}{\ensuremath{Z_{V^4_{cb}}}}
\newcommand{\gDDp}{\ensuremath{g_{D^*D\pi}}}
\newcommand{\hs}{\ensuremath{\hphantom{-}}}

\preprint{FERMILAB-PUB-12/047-T}

\title{\boldmath $B_s\to D_s/B\to D$ Semileptonic Form-Factor Ratios and \\
Their Application to BR$(B^0_s\to\mu^+\mu^-)$}

\author{Jon A.~Bailey}
\affiliation{Department of Physics and Astronomy, Seoul National University, Seoul, South Korea}

\author{A.~Bazavov}
\affiliation{Physics Department, Brookhaven National Laboratory, Upton, NY, USA}

\author{C.~Bernard}
\affiliation{Department of Physics, Washington University, St.~Louis, Missouri, USA}

\author{C.M.~Bouchard}
\affiliation{Physics Department, University of Illinois, Urbana, Illinois, USA}
\affiliation{Fermi National Accelerator Laboratory, Batavia, Illinois, USA}
\affiliation{Department of Physics, The Ohio State University, Columbus, Ohio, USA}

\author{C.~DeTar}
\affiliation{Physics Department, University of Utah, Salt Lake City, Utah, USA}

\author{Daping~Du}
\email{ddu@illinois.edu}
\affiliation{Department of Physics and Astronomy, University of Iowa, Iowa City, Iowa, USA}
\affiliation{Fermi National Accelerator Laboratory, Batavia, Illinois, USA}
\affiliation{Physics Department, University of Illinois, Urbana, Illinois, USA}

\author{A.X.~El-Khadra}
\affiliation{Physics Department, University of Illinois, Urbana, Illinois, USA}

\author{J.~Foley}
\affiliation{Physics Department, University of Utah, Salt Lake City, Utah, USA}

\author{E.D.~Freeland}
\affiliation{Physics Department, University of Illinois, Urbana, Illinois, USA}
\affiliation{Department of Physics, Benedictine University, Lisle, Illinois, USA}

\author{E.~G\'amiz}
\affiliation{Fermi National Accelerator Laboratory, Batavia, Illinois, USA}
\affiliation{CAFPE and Departamento de F\`isica Te\'orica y del Cosmos, Universidad de Granada,
Granada, Spain}

\author{Steven~Gottlieb}
\affiliation{Department of Physics, Indiana University, Bloomington, Indiana, USA}

\author{U.M.~Heller}
\affiliation{American Physical Society, Ridge, New York, USA}

\author{Jongjeong Kim}
\affiliation{Department of Physics, University of Arizona, Tucson, Arizona, USA}

\author{A.S.~Kronfeld}
\affiliation{Fermi National Accelerator Laboratory, Batavia, Illinois, USA}

\author{J.~Laiho}
\affiliation{SUPA, School of Physics and Astronomy, University of Glasgow, Glasgow, UK}

\author{L.~Levkova}
\affiliation{Physics Department, University of Utah, Salt Lake City, Utah, USA}

\author{P.B.~Mackenzie}
\affiliation{Fermi National Accelerator Laboratory, Batavia, Illinois, USA}

\author{Y.~Meurice}
\affiliation{Department of Physics and Astronomy, University of Iowa, Iowa City, Iowa, USA}

\author{E.~Neil}
\affiliation{Fermi National Accelerator Laboratory, Batavia, Illinois, USA}

\author{M.B.~Oktay}
\affiliation{Physics Department, University of Utah, Salt Lake City, Utah, USA}

\author{Si-Wei Qiu}
\affiliation{Physics Department, University of Utah, Salt Lake City, Utah, USA}

\author{J.N.~Simone}
\affiliation{Fermi National Accelerator Laboratory, Batavia, Illinois, USA}

\author{R.~Sugar}
\affiliation{Department of Physics, University of California, Santa Barbara, California, USA}

\author{D.~Toussaint}
\affiliation{Department of Physics, University of Arizona, Tucson, Arizona, USA}

\author{R.S.~Van~de~Water}
\affiliation{Physics Department, Brookhaven National Laboratory, Upton, NY, USA}

\author{Ran Zhou}
\affiliation{Department of Physics, Indiana University, Bloomington, Indiana, USA}

\collaboration{Fermilab Lattice and MILC Collaborations}
\noaffiliation
\date{\today}

\begin{abstract}
We calculate form-factor ratios between the semileptonic decays $\bar{B}^0 \to D^+\ell^-\bar{\nu}$ and 
$\bar{B}^0_{s} \to D^+_{s}\ell^-\bar{\nu}$ with lattice QCD. 
These ratios are a key theoretical input in a new strategy to determine the fragmentation fractions of 
neutral $B$ decays, which are needed for measurements of $\mathrm{BR}(B^0_s\to \mu^+\mu^-)$. 
We use the MILC ensembles of gauge configurations with 2+1 flavors of sea quarks 
at two lattice spacings of approximately 0.12~fm and 0.09~fm. 
We use the model-independent 
$z$~parametrization to extrapolate our simulation results at small recoil toward maximum recoil.
Our results for the form-factor ratios are
    $f_0^{(s)}(M^2_\pi)/f_0^{(d)}(M^2_K)  =1.046(44)_{\textrm{stat.}}(15)_{\textrm{syst.}}$ and 
    $f_0^{(s)}(M^2_\pi)/f_0^{(d)}(M^2_\pi)=1.054(47)_{\textrm{stat.}}(17)_{\textrm{syst.}}$. 
In contrast to a QCD sum-rule calculation, no significant departure from $U$-spin ($d\leftrightarrow s$) 
symmetry is observed.
\end{abstract}


\maketitle


\section{Introduction}
\label{sec:intro}

\begin{figure}[bp]
    \centering
    \includegraphics[width=4in]{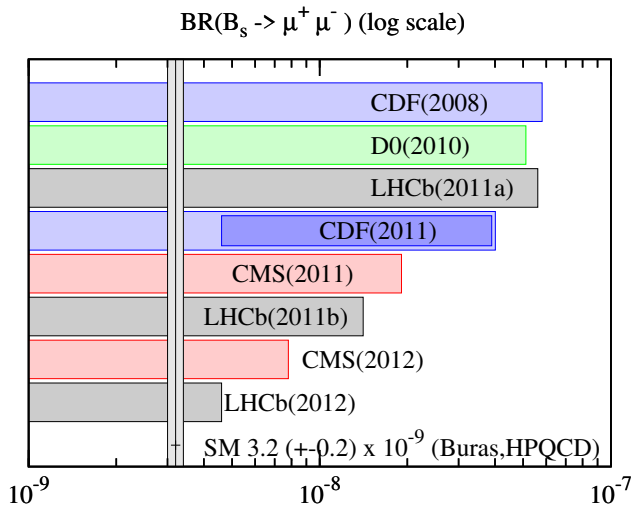}
    \caption{Comparison of the (most recent) measurements from 
        CDF~\cite{:2007kv,Aaltonen:2011fi}, 
        D\O~\cite{Abazov:2010fs}, 
        CMS~\cite{Chatrchyan:2011kr, Chatrchyan:2012rg}, 
        and LHC$b$~\cite{2011arXiv1103.2465T,:2011zq, Aaij:2012ac} with the 
        SM prediction \cite{Buras:2003td,Gamiz:2009ku} shown as a vertical band.
        The filled bars show the measured bounds of the branching ratio with a 95\% confidence.
        In the fourth bar, the inner box shows the two-sided 90\% bound from CDF~\cite{Aaltonen:2011fi}.
        Two results from the LHC$b$ in 2011 are distinguished as ``2011a'' \cite{2011arXiv1103.2465T}
        and ``2011b'' \cite{:2011zq}.}
    \label{fig:experiment_br}
\end{figure}

Recently there has been increasing interest in the rare decay $B_s^0\to\mu^+\mu^-$ which, as a 
flavor-changing neutral-current process, is forbidden at tree level in the standard model (SM). 
At the loop level, it can be mediated by weak bosons through penguin or box diagrams.
With a nonperturbative (lattice-QCD) calculation of the bag parameter $B_{B_s}$, the branching fraction has 
been predicted to be~\cite{Buras:2003td,Gamiz:2009ku} 
\be \label{eq:SM_branching}
    \textrm{BR}(B_s^0 \to \mu^+\mu^-) = 3.2(2)\times 10^{-9}.
\ee  
Several new physics models would enhance the decay rate~\cite{Buras:2010mh,Golowich:2011cx,%
Buras:2011fz,Akeroyd:2011kd,Altmannshofer:2011iv}, and, hence, observation of this process could potentially
reveal physics beyond the~SM.
Recently, several experiments~\cite{:2007kv,Abazov:2010fs,2011arXiv1103.2465T, Aaltonen:2011fi,%
Chatrchyan:2011kr,:2011zq, Chatrchyan:2012rg, Aaij:2012ac}
have published upper limits on this branching fraction, which we have compiled in 
Fig.~\ref{fig:experiment_br}. 
Moreover, CDF~\cite{Aaltonen:2011fi} reports an excess such that $\textrm{BR}(B_s^0 \to
\mu^+\mu^-)=18^{+11}_{-~9}\times10^{-9}$ or a two-sided 90\% confidence interval,
$4.6\times10^{-9}<\textrm{BR}(B_s^0 \to \mu^+\mu^-)<39\times10^{-9}$, lying above the SM prediction,
Eq.~(\ref{eq:SM_branching}).
CMS and LHC$b$, however, set upper limits that restrict the CDF region.
As statistics accumulate, especially at LHC$b$, a definitive measurement at the SM rate or higher seems
likely soon.

At a hadron collider, the extraction of $\textrm{BR}(B_s^0 \to \mu^+\mu^-)$ relies on normalization channels
such as $B_u^+\to J/\psi K^+$, $B^0_d\to K^+\pi^-$ and $B^0_s\to J/\psi \phi$ \cite{:2009ny}, through 
relations of the form 
\be
    \mathrm{BR}(B_s^0 \to \mu^+\mu^-) = \mathrm{BR}(B_q\to X) \frac{f_q}{f_s} 
        \frac{\epsilon_X}{\epsilon_{\mu\mu}} \frac{N_{\mu\mu}}{N_X},
    \label{eq:BRfsfd}
\ee
where $\epsilon$ and $N$ are, respectively, the detector efficiencies and the numbers of events.
The fragmentation fractions $f_q$ ($q=u,d,s$ or $\Lambda$) denote the probability that a $b$ quark hadronizes
into a $B_q$ meson or a $\Lambda_b$ baryon.
The fragmentation fractions $f_q$ may depend on the environment, so they are best measured \emph{in situ} in
each experiment.
Thus, improving the determination of the fragmentation ratio $f_s/f_d$ will tighten the limits and increase
the significance of measurements.

The quantity $f_s/f_d$ has generally been determined from semileptonic decays \cite{Nakamura:2010zzi}, an
approach that LHC$b$ has newly refined~\cite{Aaij:2011jp}.
Recently, Fleischer, Serra, and Tuning proposed two approaches based on measuring the ratio relative to
nonleptonic decays $\mathrm{BR}(\bar{B}^0_s\to D_s^+\pi^-)/\mathrm{BR}(\bar{B}^0\to D^+K^-)$
\cite{Fleischer:2010ay} or $\mathrm{BR}(\bar{B}^0_s\to D_s^+\pi^-)/\mathrm{BR}(\bar{B}^0\to D^+\pi^-)$
\cite{Fleischer:2010ca}.
An important ingredient in both approaches is the approximate factorization of the nonleptonic decay
amplitudes, which relies on the corrections to naive factorization of the light meson in the final state
being small and calculable \cite{Beneke:2000ry}.
The $D^+K^-$ method is favored in this regard, because it receives contributions only from color-allowed
tree-diagram-like topologies which yield smaller nonfactorizable effects~\cite{Fleischer:2010ca}.

The ratio $\mathrm{BR}(\bar{B}^0_s\to D_s^+\pi^-)/\mathrm{BR}(\bar{B}^0\to D^+K^-)$ is related to $f_s/f_d$
by analogy with Eq.~(\ref{eq:BRfsfd}).
Via factorization, the amplitudes for these nonleptonic processes can be expressed as a product of the 
light-meson decay constant and a semileptonic form factor for $B_{(s)}\to D_{(s)}\ell\nu$.
This leads to a way to measure $f_s/f_d$ \cite{Fleischer:2010ay,2011arXiv1106.4435L}:
\be \label{eq:fs_fd}
    \frac{f_s}{f_d} = 0.0743 \times \frac{\tau_{B^0}}{\tau_{B_s^0}} \times 
        \left[ \frac{\epsilon_{DK}}{\epsilon_{D_s\pi}}\frac{N_{D_s\pi}}{N_{DK}}\right ]\times
        \frac{1}{\mathcal{N}_a\mathcal{N}_F}
\ee
where $\tau$ denotes lifetimes, and the number 0.0743 is a product of ratios of well-known quantities such as the light-meson decay constants, Cabibbo-Kobayashi-Maskawa (CKM) matrix 
elements and kinematic factors.
The factorization is parametrized by~\cite{Fleischer:2010ay}
\ben
    \mathcal{N}_a & = & \left[\frac{a_1^{(s)}(D_s^+\pi^-)}{a_1^{(d)}(D^+ K^-)}\right ]^2, \label{eq:Na} \\
    \mathcal{N}_F & = & \left[\frac{f_0^{(s)}(M_\pi^2)}{f_0^{(d)}(M_K^2)} \right ]^2. \label{eq:Nf}
\een
where $a^{(q)}$ is a factor accounting for the deviation from the naive factorization and $f_0(q^2)$ is a 
form factor for the corresponding semileptonic decay. 

The hadronic method relies on theoretical inputs for $\mathcal{N}_a$ and $\mathcal{N}_F$.
In the limit of exact $U$-spin symmetry (namely the exchange of $s$ and $d$ quarks throughout the process), 
both reduce to~1. 
Fleischer, Serra, and Tuning expect the $U$-spin breaking $|\mathcal{N}_a-1|$ 
``to be at most a few percent''~\cite{Beneke:2000ry,Fleischer:2010ca}.
Based on an estimate from QCD sum rules \cite{Blasi:1993fi}, they quote either
$\mathcal{N}_F=1.3 \pm 0.1$~\cite{Fleischer:2010ay} or
$\mathcal{N}_F=1.24\pm0.08$~\cite{Fleischer:2010ca}, 
the latter of which LHC$b$ uses~\cite{2011arXiv1106.4435L}.
In either case, the biggest limitation is from the form-factor ratio $\mathcal{N}_F$.

A relation between $f_s/f_d$ and $\mathrm{BR}(\bar{B}^0_s\to D_s^+\pi^-)/\mathrm{BR}(\bar{B}^0\to D^+\pi^-)$
is derived along similar lines~\cite{Fleischer:2010ca}.
In that case, the form-factor ratio becomes $[f_0^{(s)}(M_\pi^2)/f_0^{(d)}(M_\pi^2)]^2$, i.e., with both
numerator and denominator evaluated at $q^2=M_\pi^2$.

In this paper, we calculate these two form-factor ratios using lattice QCD with 2+1 flavors of sea quarks.
We use the same set of MILC ensembles of gauge configurations~\cite{Bazavov:2009bb} and the same sequence of
bootstrap copies for both of the $B^0_s$ and $B^0$ processes, which reduces the statistical error by
correctly accounting for correlations.
We include the contributions of the first radially excited states in the fits of correlation functions to
avoid the respective systematic errors.
Such a treatment turns out to be necessary for calculations at nonzero recoil.
By fitting the correlation functions in a simultaneous and mutually constrained manner, we are able to
extract the form factors at small recoil.
We then extrapolate our lattice results to the continuum limit and to physical quark masses with the guide 
of chiral perturbation theory.
Finally, we use the model-independent $z$~parametrization~\cite{deRafael:1993ib} to extend the form factors 
toward large recoil.

We finally arrive at the result
\be
    \frac{f_0^{(s)}(M_\pi^2)}{f_0^{(d)}(M_K^2)} = 1.046 (44)(15),
    \label{eq:D-K+result} 
\ee
where the first error is statistical and the second reflects the systematic errors added in quadrature.
(Due to refinements in the analysis, Eq.~(\ref{eq:D-K+result}) differs slightly from our preliminary
result~\cite{Du:2011wn}.) 
We do not observe a large $U$-spin breaking effect.
Such a small difference between the $B_s^0$ and $B^0$ form factors is in accord, however, with recent
lattice-QCD calculations on lighter mesons like $D_{(s)}\to \pi(K)\ell\nu$ \cite{Koponen:2011ev}.
It is also in agreement with a result from heavy-meson chiral perturbation theory~\cite{Jenkins:1992qv}.

The factorization analysis of
$\mathrm{BR}(\bar{B}^0_s\to D_s^+\pi^-)/\mathrm{BR}(\bar{B}^0\to D^+\pi^-)$ is somewhat 
more complicated because of additional topologies in the decay $\bar{B}^0\to D^+\pi^-$.
A similar form-factor ratio is needed and, simply by adjusting $q^2$ in the denominator, we find
\be
    \frac{f^{(s)}_0(M_\pi^2)}{f^{(d)}_0(M_\pi^2)} = 1.054 (47)(17).
    \label{eq:D-pi+result} 
\ee
We discuss the implications of our results (\ref{eq:D-K+result}) and~(\ref{eq:D-pi+result}) in 
Sec.~\ref{sec:con}.
Here we only note that both yield fragmentation-fraction ratios $f_s/f_d$ in agreement with 
LHC$b$'s recent measurement via semileptonic methods~\cite{Aaij:2011jp}.

This paper is organized as follows.
In Sec.~\ref{sec:ff}, we summarize the formalism and our strategy to extract the form factors at nonzero
recoil.
We provide simulation details in Sec.~\ref{sec:sims}.
We describe the methodology used to extract the form factors from the two- and three-point correlation
functions with the given gauge configurations.
This fitting procedure is crucial to our analysis.
In Sec.~\ref{sec:chiral} we describe the chiral-continuum extrapolation using the corresponding chiral
perturbation theory.
In Sec.~\ref{sec:z}, these results are then extrapolated to the region of small momentum transfer using a
model-independent parametrization.
We also compare here a related form factor, which we obtain as a by-product, with the experimental results.
In Sec.~\ref{sec:sys}, we account for the systematic errors that arise in our analysis and present a full
error budget.
Finally, in Sec.~\ref{sec:con}, we present our results, compare with previous results and discuss prospects
and connections to current and future experiments.
The Appendix specifies the functional form of the chiral extrapolation in detail.


\section{Semileptonic \boldmath $B_{(s)}\to D_{(s)}\ell\nu$ form factors from lattice QCD}
\label{sec:ff}

The hadronic matrix elements of the semileptonic decays $B_{(s)}\to D_{(s)}\ell\nu$ can be parametrized by 
\be
    \langle D(p') | \mathcal{V}^\mu | B(p) \rangle = 
        f_+(q^2) \left[ (p+p')^\mu - \frac{M_B^2-M_D^2}{q^2}q^\mu \right] + 
        f_0(q^2) \frac{M_B^2-M_D^2}{q^2} q^\mu,
\ee
where $q=p-p'$ is the momentum transfer and $\mathcal{V}^\mu = \bar{c}\gamma^\mu b$ is the (continuum)
\pagebreak
vector current.
Another parametrization uses velocity 4-vectors $v = p/M$ instead of momentum $p$~\cite{Neubert:1993mb}, 
\be
    \frac{\langle D(p') | \mathcal{V}^\mu | B(p) \rangle }{ \sqrt{M_B M_D}} =
        h_+(w)(v + v')^\mu  + h_-(w) (v-v')^\mu, 
\ee
where $w=v\cdot v'=(M_B^2+M_D^2-q^2)/2M_BM_D$ describes the recoil of the process. 
The $h_\pm$ parametrization is convenient for lattice QCD, both for numerical
simulation~\cite{Hashimoto:1999yp} and for matching lattice gauge theory to
continuum~QCD~\cite{Kronfeld:2000ck,Harada:2001fi}.

The lattice-QCD calculation of $h_+$ in the zero-recoil limit has been investigated using double
ratios~\cite{Hashimoto:1999yp}, 
\be
    R_+ = \frac{\langle D|\bar{c}\gamma^0b|B\rangle \langle B|\bar{b}\gamma^0c|D\rangle}%
        {\langle D|\bar{c}\gamma^0c|D\rangle \langle B|\bar{b}\gamma^0b|B\rangle} = 
        |h_+(1)|^2
    \label{eq:R}
\ee
with all states at rest.
To proceed analogously at nonzero momentum, we introduce the following single ratios:
\ben
    \bm{a} & \equiv & \frac{\langle D(\bm{p}) | \bar{c}\bm{\gamma} b | B(\bm{0}) \rangle}
        {\langle D(\bm{0}) | \bar{c}\gamma^0 b | B(\bm{0}) \rangle} =
        \frac{h_+(w)-h_-(w)}{2h_+(1)}\bm{v},
    \label{eq:ai} \\
    \bm{b} & \equiv & \frac{\langle D(\bm{p}) | \bar{c}\bm{\gamma} b| B(\bm{0}) \rangle}
        {\langle D(\bm{p}) | \bar{c}\gamma^0 b | B(\bm{0}) \rangle} =
        \frac{h_+(w)-h_-(w)}{(w+1)h_+(w)-(w-1)h_-(w)}\bm{v},
    \label{eq:bi} \\
    \bm{d} & \equiv & \frac{\langle D(\bm{p}) | \bar{c}\bm{\gamma} c| D(\bm{0}) \rangle}
        {\langle D(\bm{p}) | \bar{c}\gamma^0 c | D(\bm{0}) \rangle} = \frac{\bm{v}}{1+w},
    \label{eq:di}
\een 
where the last follows from vector current conservation, $h_-^{D\to D}(w)=0$.

We can write down the equations that manifest the relations between the ratios and the form factors
\ben
    w & = & \frac{1+\bm{d}\cdot\bm{d}}{1-\bm{d}\cdot\bm{d}}, 
    \label{eq:wd} \\
    \frac{h_+(w)}{h_+(1)} & = &  \frac{a_i}{b_i}-\bm{a}\cdot\bm{d}, 
    \label{eq:h+abd} \\
    \frac{h_-(w)}{h_+(1)} & = &  \frac{a_i}{b_i}-\frac{a_i}{d_i}.
    \label{eq:h-abd} 
\een
In Eq.~(\ref{eq:ai}), we have a ratio $\bm{a}$ between matrix elements involving a final $D$ meson
with nonzero and zero spatial momentum.
The purpose is to make use of the correlations in the uncertainties between the two.
The form factor at zero recoil can be extracted precisely via $R_+$~\cite{Hashimoto:1999yp}, and we
find that these ratios aid calculations at nonzero recoil in a similar way.

With $h_\pm(w)$ in hand, one can obtain the form factors $f_+(q^2)$ and $f_0(q^2)$,
\ben
    f_+(q^2) & = & \frac{1}{2\sqrt{r}} \left[ (1+r) h_+(w) - (1-r) h_-(w) \right] ,
    \label{eq:f+} \\
    f_0(q^2) & = & \sqrt{r} \left [ \frac{w+1}{1+r} h_+(w) - \frac{w-1}{1-r} h_-(w) \right ], 
    \label{eq:f0} 
\een
where $r=M_D/M_B$ and $q^2=M_B^2+M_D^2-2wM_BM_D$.
Equations~(\ref{eq:h+abd}) and~(\ref{eq:h-abd}) both contain the factor $h_+(1)$, so we write  
\be
    f_0(q^2) = h_+(1) \tilde{f}_0\left(w(q^2)\right).
\ee
In the formulas until now, we have not specified the spectator mass, 
\pagebreak
so they apply to both the $B\to D$ and $B_s\to D_s$ processes.
With this notation, the desired ratio of the form factors is then
\be
    \frac{f_0^{B_s\to D_s}(M_\pi^2)}{f_0^{B\to D}(M_K^2)} = \frac{h_+^{B_s\to D_s}(1)}{h_+^{B\to D}(1)}
        \frac{ \tilde{f}_0^{B_s\to D_s}\left(w(M_\pi^2)\right)}{\tilde{f}_0^{B\to D}\left(w(M_K^2)\right)},
\label{eq:f0ratio}
\ee
where the first factor is obtained from the ratios $R_+^{B\to D}$ and $R_+^{B_s\to D_s}$ and the last from 
the expressions in Eqs.~(\ref{eq:h+abd}) and~(\ref{eq:h-abd}).

On the lattice, we define a vector current $V^\mu = \ZVcc^{1/2}\ZVbb^{1/2}\bar{\Psi}_c\gamma^\mu\Psi_b$
\cite{Hashimoto:1999yp,Harada:2001fi}, where the factors $\ZVQQ$ normalize the flavor charge.
The matching between the lattice and the continuum physics can be bridged by the relation
$\mathcal{V}^\mu=\rhoV{\mu}V^\mu$, where $\rhoV{\mu}^2=\ZVbc\ZVcb/\ZVbb\ZVcc$.
The normalization factors $\ZVQQ$ cancel in the ratios in Eqs.~(\ref{eq:R})--(\ref{eq:di}).
The factor $\rhoV{4}$ has been verified to be very close to~1 with one-loop perturbation theory with
unimproved gluons~\cite{Harada:2001fi}.
Calculations of $\rhoV{\mu}$ with improved gluons (as used here; cf.\ Sec.~\ref{sec:sims}) are in progress.
Given the ratio structure in Eq.~(\ref{eq:f0ratio}), it is clear that the (small) contributions from
$\rhoV{\mu}-1$ should largely cancel.
Thus, in this analysis, we take $\rhoV{\mu}=1$ and estimate the uncertainty from this choice in
Sec.~\ref{sec:sys}.


\section{Simulations and fitting methodology}
\label{sec:sims}

\subsection{Data Setup and the Lattice Simulations}

Our calculation uses four ensembles of MILC's (2+1)-flavor asqtad configurations~\cite{Bazavov:2009bb} at two
lattice spacings, $a\approx 0.12~\mathrm{fm}$, $0.09~\mathrm{fm}$, which we refer to as the ``coarse'' and
``fine'' lattices, respectively.
The configurations were generated with an $O(a^2)$ Symanzik improved gauge action
\cite{Weisz:1982zw,Curci:1983an,Weisz:1983bn,Luscher:1984xn}.
The coarse (fine) ensembles used here have a lattice size of $20^3\times 64$ ($28^3\times96$), so in both
cases the spatial size is $L\approx2.4~\mathrm{fm}$.
The four ensembles have different sea-quark masses, so, for the sake of convenience, we label them C020,
C007, F0062, and F0124.
Details on the parameters that we use in the simulations are summarized in Table~\ref{tab:ensembles}.
\begin{table}[tp]
\caption{\label{tab:ensembles} Parameters of the MILC asqtad ensembles of configurations and the valence
    quarks used in this analysis.}
\begin{tabular}{ccccccccc}
\hline\hline
Ensemble  & 
$a$ (fm) & 
$am_l/am_h$ & 
$ N_{\mathrm{confs}}$ & 
$\kappa_c$  &
$\kappa_b$  & 
$c_{SW}$ & 
\;$am_x(B\to D)$ & 
\;$am_x(B_s\to D_s$)\cr
 \hline 
    C020\; &$\approx$ 0.12&0.020/0.050 & 2052 & 0.1259& 0.0918 & 1.525 & 0.020 & 0.0349\cr 
    C007\; &$\approx$ 0.12&   0.007/0.050 & 2110 & 0.1254& 0.0901 & 1.530 & 0.007 & 0.0349\cr 
    F0124\; &$\approx$ 0.09&0.00124/0.031 & 1996 & 0.1277& 0.0982 & 1.473 & 0.0124 & 0.0261\cr 
    F0062\; &$\approx$ 0.09&0.0062/0.031 & 1931 & 0.1276& 0.0979 & 1.476 & 0.0062 & 0.0261\cr
\hline\hline
\end{tabular}
\end{table}
The strange and light sea quarks are simulated using the asqtad-improved staggered
action~\cite{Blum:1996uf,Orginos:1998ue,Orginos:1999cr,Lepage:1998vj,Lagae:1998pe}.
The asqtad action is also used for our strange and light valence quarks.
The heavy charm and bottom quarks are simulated using the Sheikholeslami-Wohlert (SW) clover
action~\cite{Sheikholeslami:1985ij} with the Fermilab interpretation \cite{ElKhadra:1996mp}.
We simulate the $B\to D$ and $B_s \to D_s$ decays on the same ensembles, so that correlations reduce the
statistical uncertainty in the ratios.
For the $B\to D$ decay, the valence light-quark mass is taken to be the same as the sea-quark mass, i.e., we
stick to ``full QCD'' data with $m_x=m_l$, while for the $B_s\to D_s$ process, we set the valence
strange-quark mass to be close to its physical value, $m_x=m_s$.
The charm and bottom quarks in our calculation are tuned to their physical values up to a tuning uncertainty.
Columns~2--9 in Table \ref{tab:ensembles} list, respectively, the approximate lattice spacings,
light$/$strange sea quark masses, number of configurations, the hopping parameter $\kappa_{b(c)}$, the
coefficient for the clover term $c_{\rm SW}$, and the light valence-quark masses used in the $B\to D$ and
$B_s\to D_s$ simulations.
The quark masses here are all in lattice units.

We obtain the matrix elements appearing in Sec.~\ref{sec:ff} from the following three-point correlation
functions: 
\ben
    C^{DV^\mu B}_3(0,t,T;\bm{p}) &=& \sum_{\bm{x},\bm{y}} \; \langle \mathcal{O}_D(0, \bm{0})
        \overline{\Psi}_c i\gamma^\mu \Psi_b(t, \bm{y}) \mathcal{O}^\dagger_B(T, \bm{x})  \rangle \; 
        e^{i \bm{p}\cdot\bm{y}}, 
    \label{eq:3pt:B2D} \\
    C^{DV^\mu D}_3(0,t,T;\bm{p}) &=& \sum_{\bm{x},\bm{y}} \; \langle \mathcal{O}_D(0, \bm{0})
        \overline{\Psi}_c i\gamma^\mu \Psi_c(t, \bm{y}) \mathcal{O}^\dagger_D(T, \bm{x})  \rangle \; 
        e^{i \bm{p}\cdot\bm{y}},
    \label{eq:3pt:D2D} \\
    C^{BV^4 B}_3(0,t,T;\bm{0}) &=& \sum_{\bm{x},\bm{y}} \; \langle  \mathcal{O}_B(0, \bm{0})
        \overline{\Psi}_b i\gamma^4 \Psi_b(t, \bm{y}) \mathcal{O}^\dagger_B(T, \bm{x}) \rangle ,
    \label{eq:3pt:B2B} 
\een
where the sum over $\bm{x}$ sets the $B$ meson at rest,
and the sum over $\bm{y}$ selects the final-state $D$-meson momentum.
The final $D$ meson is simulated with several small spatial momenta which are the lowest possible values for
the finite spatial volumes: $\bm{p}=2\pi(0,0,0)/L$, $2\pi(1,0,0)/L$, $2\pi(1,1,0)/L$, $2\pi(1,1,1)/L$,
$2\pi(2,0,0)/L$, and permutations.
To increase statistics, data are generated at four different source times, spaced evenly along the temporal
extent of the lattice.
The zero-momentum correlation functions for $D\to D$ and $B\to B$ serve as normalization, as discussed above.
The $D\to D$ correlation function with a nonzero final state momentum is used to extract~$w$ via
Eqs.~(\ref{eq:di}) and~(\ref{eq:wd}).

The analysis below also requires the two-point function
\begin{equation} \label{eq:2pt:XX}
    C_{2}^{X}(t,\bm{p}) = \sum_{\bm{x}} e^{i\bm{p}\cdot\bm{x}}\,
        \langle\mathcal{O}_X^\dagger(t,\bm{x})\mathcal{O}_X(0,\bm{0})\rangle,
\end{equation}
where $X=B$ or $D$.

We simulate the daughter meson with two different choices for the interpolation operator $\mathcal{O}_X$:
with a 1S-wave smearing and without any smearing \cite{Bernard:2008dn}.
The smearing optimizes the overlap of the operator with the ground-state wave function of the meson, so the
two choices have different excited-state contributions but the same energies.
For the three-point correlation functions, we always use a 1S-smearing source for the extended quark
propagator.

\subsection{From correlators to form factors}

In general, two- and three-point functions, such as those in Eqs.~(\ref{eq:3pt:B2D})--(\ref{eq:2pt:XX}), can 
be expressed as~\cite{Wingate:2002fh}, 
\ben
    C_2^{X}(t,\bm{p}) & = & \sum_{k=0} (-1)^{kt} |Z_k(\bm{p})|^2\left[ e^{-E_k(\bm{p})t} +
e^{-E_k(\bm{p})(T-t)} \right],
    \label{eq:2pt}  \\
    C_3^{YV^\mu X}(0,t,T; \bm{p}) & = & \sum_k\sum_\ell (-1)^{kt}(-1)^{\ell(T-t)} A^\mu_{k\ell}(\bm{p})
    \;e^{-E_k(\bm{p}) t}\; e^{-M_\ell (T-t)} \label{eq:3pt_general}
\een
where $E_k$ ($M_\ell$) are the energy levels of $Y$~($X$) and $A^\mu_{k\ell}$ are coefficients of the
transition $X_\ell \to Y_k$.
We use four states to fit the two-point functions in Eq.~(\ref{eq:2pt}).
We include the same set of states to fit the three-point functions and the number of states can be reduced to
two (the ground and first excited states) by using some averaging method (next paragraph).
If the time differences between the source (0) and vector current ($t$) and that between vector current and
sink ($T$) in Eq.~(\ref{eq:3pt_general}) are sufficiently large , i.e., $|t|\to\infty$ and $|T-t|\to\infty$,
only the lowest energy level will survive.
Then, we have ($B_s\to D_s$ follows similarly)
\ben
    R_+ & \leftarrow & \frac{C_3^{DV^4 B}(0,t,T; \bm{0})}{C_3^{BV^4 B}(0,t,T; \bm{0})}
        \frac{C_3^{BV^4 D}(0,t,T; \bm{0})}{C_3^{DV^4 D}(0,t,T; \bm{0})},
    \label{eq:R+_extract} \\
    a_i & \leftarrow & \frac{C_3^{DV^i B}(0,t,T; \bm{p})}{C_3^{DV^4 B}(0,t,T; \bm{0})} \left( 
        \frac{|Z_0(\bm{0})|}{|Z_0(\bm{p})|}\sqrt{\frac{E_0(\bm{p})}{E_0(\bm{0})}}
        e^{[E_0(\bm{p})-E_0(\bm{0})]t} 
        \right),
    \label{eq:ai_extract} \\
    b_i & \leftarrow & \frac{C_3^{DV^i B}(0,t,T; \bm{p})}{C_3^{DV^4 B}(0,t,T; \bm{p})},
    \label{eq:bi_extract} \\
    d_i & \leftarrow & \frac{C_3^{DV^i D}(0,t,T; \bm{p})}{C_3^{DV^4 D}(0,t,T; \bm{p})},
    \label{eq:di_extract}
\een
where $\leftarrow$ means that the left-hand side is output of an analysis procedure.
In practice, the separations between the current insertion and the source$/$sink, $t$ and $T-t$, are often
not large enough to suppress the excited states completely.
The factor inside the parentheses in Eq.~(\ref{eq:ai_extract}) cancels the leading time dependence of the 
ratio of three-point functions with different momenta; both $Z_0(\bm{p})$ and $E_0(\bm{p})$ come from 
fitting $C_2^D$ as suggested by Eq.~(\ref{eq:2pt}).
Instead of fitting for plateaus, we extract the matrix-element ratios on the left-hand sides of
Eqs.~(\ref{eq:R+_extract})--(\ref{eq:di_extract}) by fitting the right-hand sides in a way that incorporates
excited states.
\begin{figure}[bp]
    \centering
    \includegraphics[width=0.5\textwidth, trim=5mm 0 0 0]{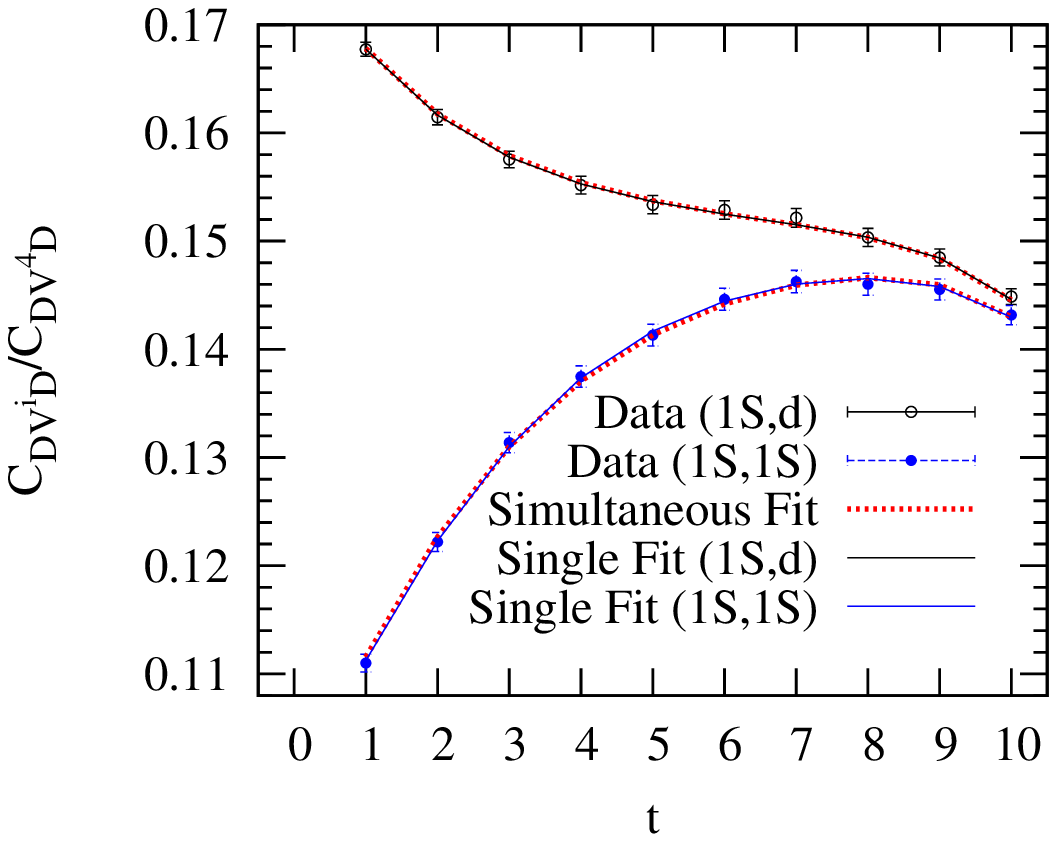}\hfill
    \includegraphics[width=0.5\textwidth, trim=5mm 0 0 0]{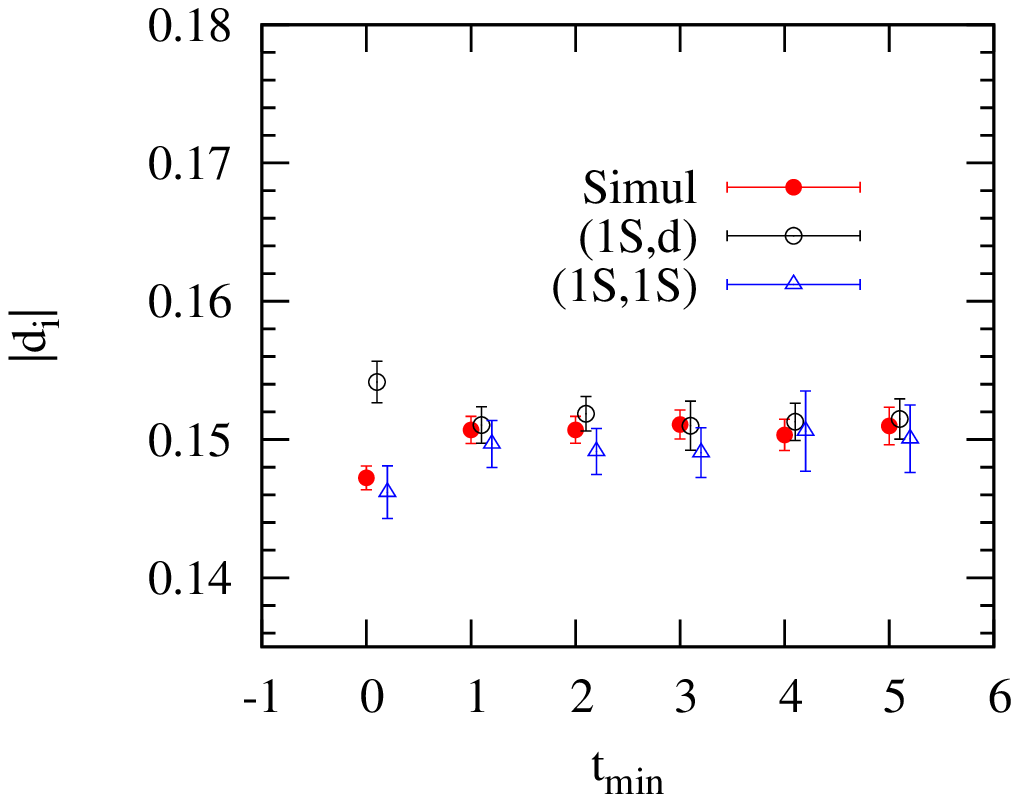}
    \caption{A sample fit of the single ratio $d_i$ with the final $D$ meson carrying momentum
        $\bm{p}=2\pi(1,0,0)/L$ on the ensemble C020.
        The left graph shows the data with 1S-smearing (1S, 1S) (filled circles, data along the bottom 
        curve) and that with partial smearing (1S, d) (open circles, data along the top curve) fitted 
        separately (solid curves) and simultaneously (dashed curves).
        The (1S, 1S), (1S, d), and simultaneous fits have $\chi^2/\text{d.o.f.}=1.0$, 1.1 and 
        0.85, respectively.
        Fit results over the interval $t\in[t_{\mathrm{min}}, 10]$ are shown in the right graph and are seen 
        to be stable for $t_{\mathrm{min}}\ge1$.}
    \label{fig:2pt_sample}
\end{figure}

We can write the first few terms in Eq.~(\ref{eq:3pt_general}) as~\cite{Bernard:2008dn}
\ben
    C^{YV^\mu X}_3(0,t,T) & = & A^\mu_{00}(t) + (-1)^t A^\mu_{10}(t) + (-1)^{T-t}A^\mu_{01}(t) + 
        (-1)^TA^\mu_{11}(t) \nonumber \\
        & + & A^\mu_{02}(t) + A^\mu_{20}(t) + \textrm{higher excitations},
\een
where $A^\mu_{k\ell}(t)\equiv A^\mu_{k\ell}\;e^{-E_kt}e^{-M_\ell(T-t)}$.
The terms $A_{10}, A_{01}, A_{11}$ are the contributions from the opposite-parity states that are introduced
by the operator involving staggered quarks.
We can reduce their effects by making use of the fact that they oscillate with either $t$ or $T$.
The contamination from $A_{10}$ and $A_{01}$ is minor because no obvious oscillation is visible with any of
the three-point functions.
Based on the plots of the ratios calculated from different source-sink separations $T$, there is no sizable
contribution from $A_{11}$ either.
With the four source times, which are evenly distributed in the temporal interval of the lattice, we separate
two of the sinks from the sources by an even number of time slices and the other two by a neighboring odd
number.
We apply the following averaging method to further reduce the effect from $A_{11}$~\cite{Bernard:2008dn}
\be
    \bar{R}(0,t,T) = \frac{1}{2} R(0,t,T) + \frac{1}{4} R(0, t, T+1) + \frac{1}{4} R(0,t+1, T+1) 
\ee
where $R$ stands for any of the correlation-function ratios corresponding to $R_+$, $a_i$, $b_i$, or $d_i$.
With this averaging procedure, the $A_{11}$ terms are suppressed by a factor of 6--10 and the $A_{01}$ and
$A_{10}$ terms, already small, are suppressed by a factor of about 2.
Hence, the systematic error arising from neglecting terms $A_{01},A_{10}$ and $A_{11}$ can be safely dropped.

The foregoing analysis enables us to use a simple fitting scheme for the ratios including only the
contributions from $A_{00}, A_{20}, A_{02}$.
At the lowest order, the functional forms for determining $a_i$, $b_i$ and $d_i$ are then
\ben
    \frac{C^{DV^iB}_{3}(0,t,T;\bm{p})}{C^{DV^4B}_{3}(0,t,T;\bm{0})} & & \hspace*{-1.2em} 
        \frac{|Z_0(\bm{0})|}{|Z_0(\bm{p})|}\sqrt{\frac{E_0(\bm{p})}{E_0(\bm{0})}}
            e^{[E_0(\bm{p})-E_0(\bm{0})]t} \nonumber \\
        & = & a_i
        \left[1 + \mathscr{A}_{02}\;e^{- \Delta M(T-t)} + \mathscr{A}_{20}\;e^{- \Delta E(\bm{p}) t} +
            \mathscr{A}'_{20}\;e^{- \Delta E(\bm{0}) t}\right] e^{\delta t} ,
    \label{eq:ai_fit} \\
    \frac{C^{DV^iB}_{3}(0,t,T;\bm{p})}{C^{DV^4B}_{3}(0,t,T;\bm{p})} & = & b_i
        \left[1 + \mathscr{B}_{02}\;e^{- \Delta M(T-t)} + \mathscr{B}_{20}\;e^{- \Delta E(\bm{p}) t}\right],
    \label{eq:bi_fit} \\
    \frac{C^{DV^iD}_{3}(0,t,T;\bm{p})}{C^{DV^4D}_{3}(0,t,T;\bm{p})} & = & d_i
        \left[1 + \mathscr{D}_{02}\;e^{-\Delta E(\bm{0})(T-t)} + 
        \mathscr{D}_{20}\;e^{-\Delta E(\bm{p}) t}\right].
    \label{eq:di_fit}
\een  
The fit parameter $\delta$ in Eq.~(\ref{eq:ai_fit}) allows for imperfect cancellation of the leading $t$ 
dependence.
The parameters $\Delta E(\bm{p})=E_2(\bm{p})-E_0(\bm{p})$ ($\Delta M=M_2-M_0$) denote the splittings between
the $D$-meson energy ($B$-meson mass) and its first radial excitation.
We find that the double ratio $R_+$ is so weakly affected by excited states that it suffices to fit it to a 
constant in~$t$.
Adding terms to describe excited states in $R_+$ leads to changes no bigger than the statistical
errors.
The energy splittings $\Delta E(\bm{p})$, $\Delta M$ in these expressions are also constrained by the
two-point functions, Eq.~(\ref{eq:2pt}).
\begin{figure}[bp]
    \centering
    \includegraphics[width=0.3\textwidth,trim=48pt 8pt 44pt 8pt]{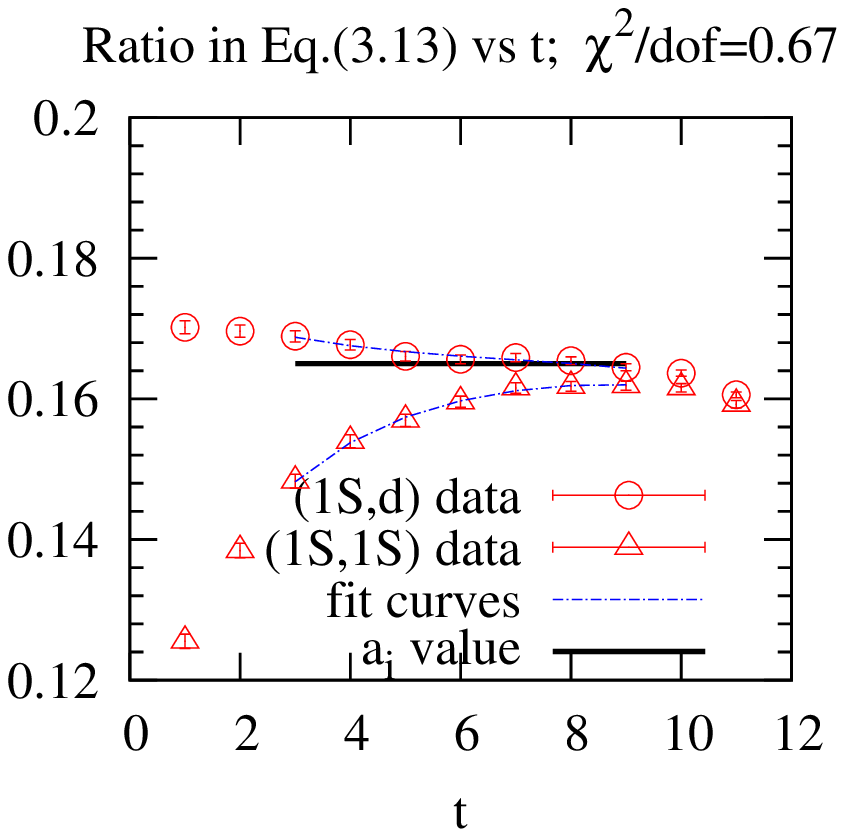}\hfill
    \includegraphics[width=0.3\textwidth,trim=48pt 8pt 44pt 8pt]{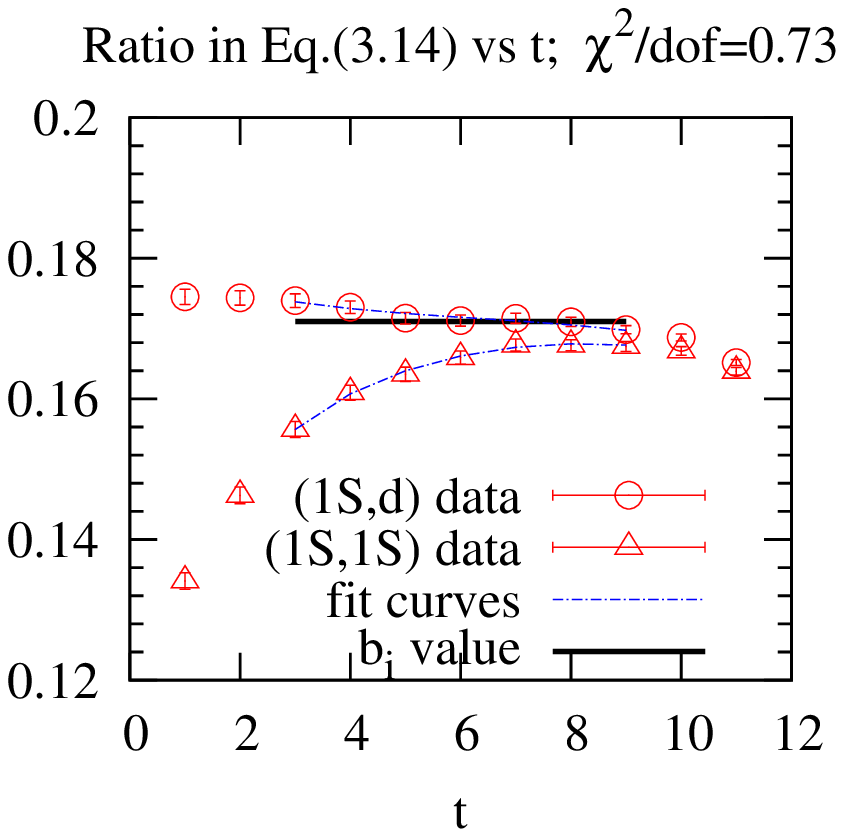}\hfill
    \includegraphics[width=0.3\textwidth,trim=48pt 8pt 44pt 8pt]{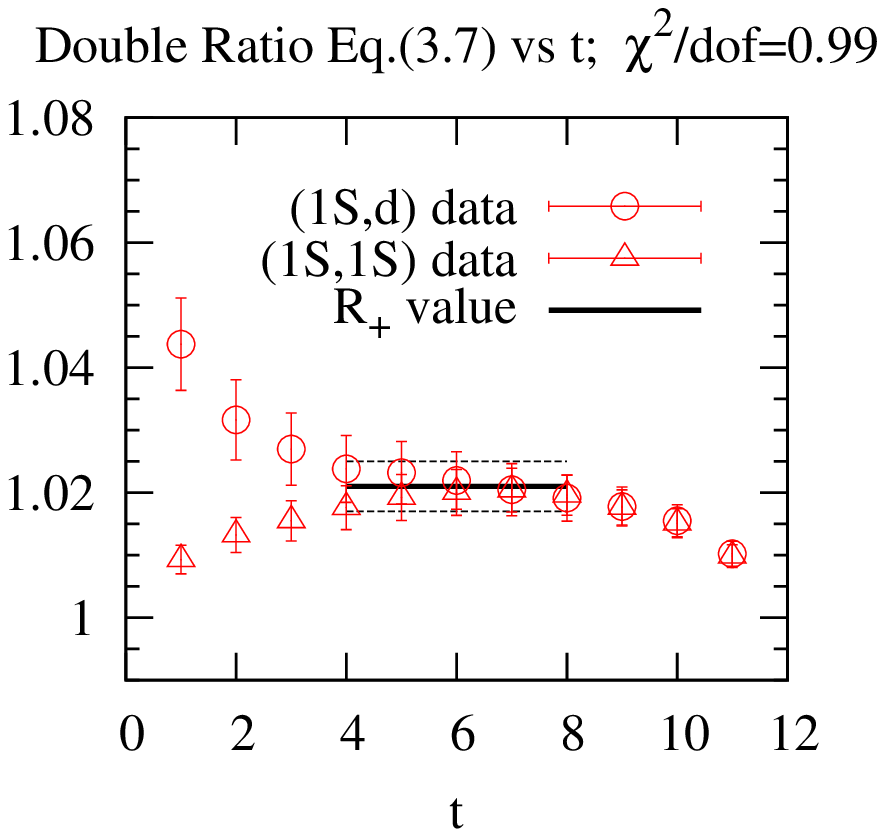}
    \caption{Sample fits of the ratios $a_i$, $b_i$, and $R_+$ from the ensemble C020. 
        The triangles and circles correspond to the data with $1S$ smearing (1S,1S) and partial smearing 
        (1S,d), respectively. 
        In the first two graphs, the final $D$ meson carries a spatial momentum $\bm{p} = 2\pi(1,0,0)/L$. 
        The dot-dashed curves indicate the best fits and they are in good agreement with the data points. 
        The third graph shows the fit for the double ratio $R_+$. 
        The horizontal lines in these graphs show the resulting values of $a_i$, $b_i$, and $R_+$, as well 
        as the ranges included in the fits.} 
    \label{fig:sr_sample}
\end{figure}

We employ two fit procedures to determine the ratios.
One, which we explain first, is simpler.
The other is more complicated but yields better results, as we explain below, so we take it as our primary
analysis and use the simple method as a cross check.

\subsection{Two-step Fit}
\label{sec:2step}

The simpler method proceeds in two steps.
We first fit the two-point functions to obtain the energies, energy splittings, and overlaps: $E(\bm{p})$,
$\Delta E(\bm{p})$, $\Delta M$, and $Z(\bm{p})$.
We use constrained curve fitting and priors \cite{Lepage:2001ym, Morningstar:2001je}.
The priors in these fits impose no real constraint.
Second, we take the energy splittings from these two-point fits as priors when fitting the ratios of
three-point functions.
The priors for the amplitudes on the right-hand sides of Eqs.~(\ref{eq:ai_fit})--(\ref{eq:di_fit}) are again
taken wide enough to impose no real constraint.
Below we call this approach the ``two-step fit.''

As discussed above, we have data for smeared and local interpolating operators.
These different correlator ratios have different excited-state contributions.
To determine the matrix-element ratios, we fit the correlation-function ratios of the two smearing types
either separately or simultaneously, as illustrated in Fig.~\ref{fig:2pt_sample}.
Although the two smearing types follow rather different curves, they arrive at consistent values of $a_i$,
$b_i$, and $d_i$ as described in Eqs.~(\ref{eq:ai_fit})--(\ref{eq:di_fit}).
Figure~\ref{fig:2pt_sample} shows that the separate fit and the simultaneous fit give consistent results for
the ratio~$d_i$.
The other two single ratios $a_i$, $b_i$ can be determined in a similar way, which can be seen in the sample
fit in Fig.~\ref{fig:sr_sample}.
\begin{figure}[b]
    \centering
    \vspace*{-24pt}
    \includegraphics[width=4in]{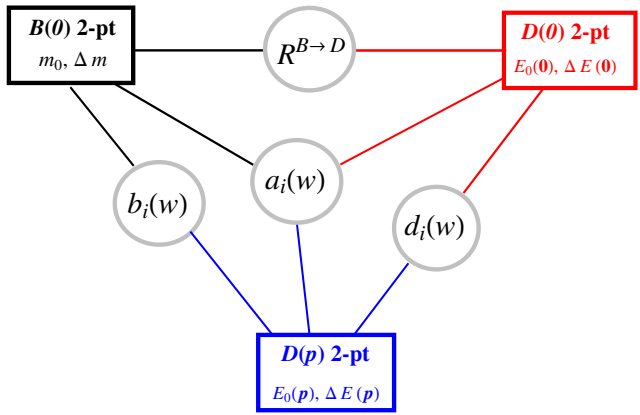}
    \vspace*{-36pt}
    \caption{Diagram showing which correlators and correlator ratios influence energies and amplitude ratios.
         Energy splittings of the initial and final mesons $B(\bm{0})$, $D(\bm{0})$, and $D(\bm{p})$ 
        are determined from the two-point functions (boxes) as well as the ratios (circles).
        The lines connecting them indicate their common dependence on the splittings.
        Altogether seven correlation functions are included in the combined fit.}
    \label{fig:combine_diagram}
\end{figure}

\vspace*{-7pt}
\subsection{Combined Fit}

Our preferred fit treats the two-point functions and three-point-function ratios simultaneously.
This approach allows all correlation functions to influence the output energies and overlaps, combining all 
information at hand, including the correlations between the two- and various three-point functions. 
We refer to this procedure as our ``combined fit'', in contrast with the two-step fit described above.
In particular, all correlation functions are then treated on the same footing in the determination of the 
energy splittings. 
Figure~\ref{fig:combine_diagram} shows the relationships between the ratios and two-point functions,
building up constraints among correlation functions of zero and nonzero final momenta. 
The single ratios $a_i$ also require the two-point $Z(\bm{p})$ factors from the relevant three two-point
functions.

\begin{figure}
    \centering
    \includegraphics[width=0.48\textwidth, trim=24pt 0 24pt 0]{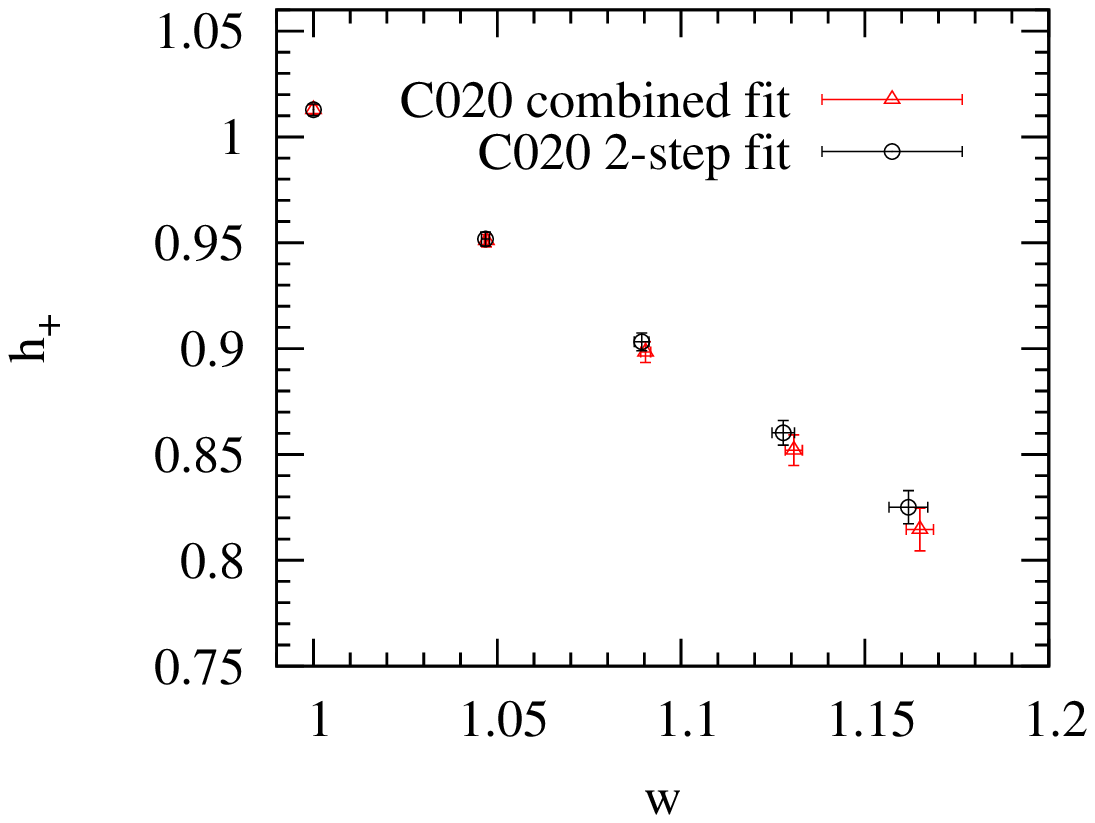}
    \includegraphics[width=0.48\textwidth, trim=24pt 0 24pt 0]{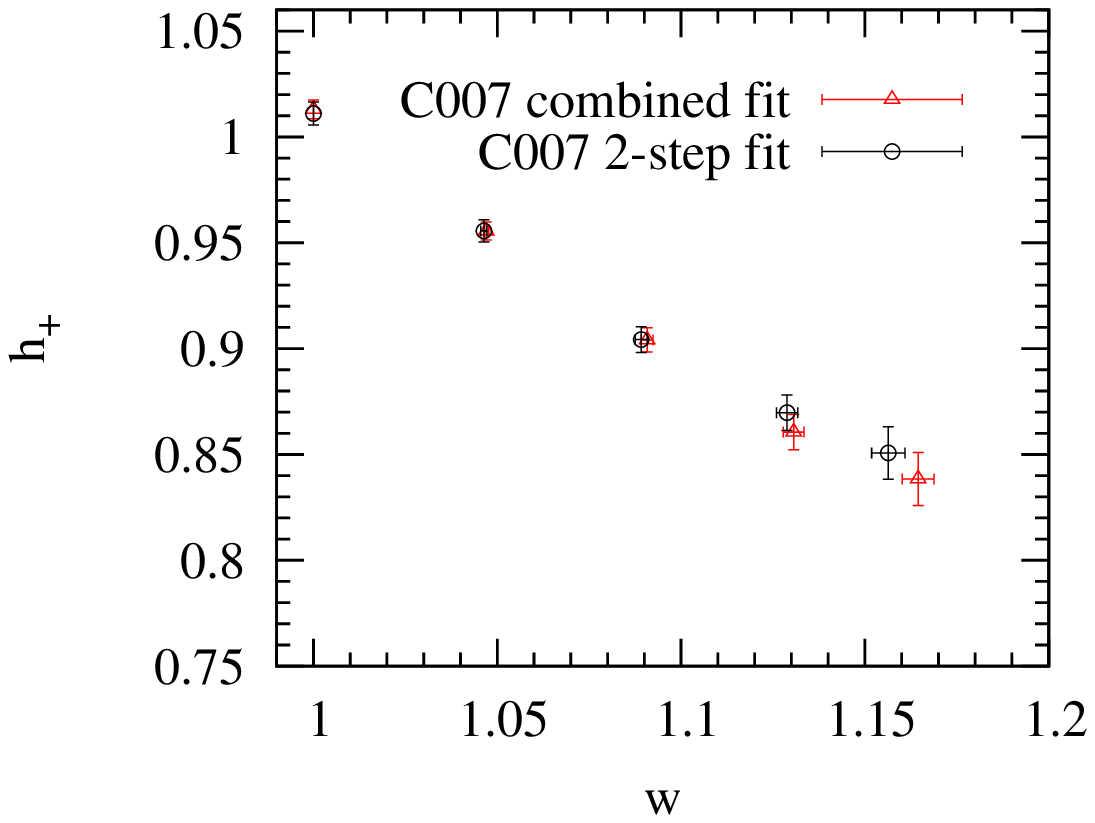}
    \caption{Comparison of $h_+(w)$ obtained from the two-step fit and combined fit for the two coarse
        ensembles C020 (left) and C007 (right).}
    \label{fig:hpcompare}
\end{figure}

The combined fit is repeated for each value of the momentum. 
The resulting single ratios $a_i$, $b_i$, and $d_i$ and the double ratio $R_+$ at the corresponding 
recoil are then used to compute the form-factor ratios $h_\pm(w)/h_+(1)$ and $h_+(1)$.
To verify the results from the two approaches, we compare the resulting $h_+(w)$ of the two coarse ensembles 
C007, C020 in Fig.~\ref{fig:hpcompare}. 
In general, they are in very good agreement at the five values of $w$ where we have data, while the 
combined fit has slightly better precision at small recoil.

The combined fit turns out to have other more important advantages over the two-step fit.
First, the resulting $h_\pm(w)/h_+(1)$ is more stable with the combined fit than the two-step fit, where the
fitting range must be determined carefully.
This stability stems from the fact that the combined fit does a better job of resolving the correlated
statistical fluctuations at zero and nonzero recoil.
Second, the combined fit helps to reduce the systematic error due to excited states.
We account for the excited state contribution in the fit for single ratios to work around the fact that
sink-source separations cannot be taken satisfactorily large.
Although the towers of the excited states are not the focus of this paper, the proper accounting of their
contributions is important, because they influence ratio results.
The combined fit procedure resolves the excited states more stably than the two-step fit.
Third, the resulting form factors $h_\pm(w)$ using the combined fit at these values of $w$ are more 
consistent with each other than those using the two-step fit.
This can be seen when one attempts to fit the results at different recoil to a chiral effective theory.
Section~\ref{sec:chiral} shows that the combined fit procedure results in a more reliable chiral
extrapolation than the two-step fit, and hence we use it throughout our analysis.
That said, as shown in Fig.~\ref{fig:hpcompare} and Table~\ref{tab:hpcompare}, the two fitting procedures 
give seemingly good pointwise consistency.

\begin{table}[tp]
\caption{Comparison of the results of $w$ [from Eq.~(\ref{eq:wd})] and $h_+$
[from Eqs.~(\ref{eq:R}) and~(\ref{eq:h+abd})] using 2-step fit and combined fit procedures.}
\label{tab:hpcompare}
\begin{tabular}{cccccc}
\hline\hline
momentum $(2\pi/L)$ &  &
C007 (2-step) & 
C007 (combined) & 
C020 (2-step) & 
C020 (combined)  \cr
 \hline 
(0,0,0) &  $w$  &  1&  1& 1 & 1 \cr
      &  $h_+$ & 1.011(6) & 1.013(4) & 1.013(3) & 1.013(2) \cr
(1,0,0) & $w$   &  1.0464(9) & 1.0470(7) & 1.0468(10) & 1.0470(6) \cr
    & $h_+$ &  0.956(5)  & 0.956(4)  & 0.952(3) & 0.951 (3) \cr
(1,1,0) & $w$  &  1.089(2)  & 1.091(2) & 1.089(2)  & 1.090(1) \cr
    & $h_+$ &  0.904(6)  & 0.904(4) & 0.903(4)  &  0.898(5) \cr
(1,1,1) & $w$ &   1.129(3)  & 1.131(3) & 1.128(3) & 1.131(2) \cr
    & $h_+$ & 0.870(8) &  0.861(6) & 0.860(6) & 0.852(7) \cr
(2,0,0) & $w$ &  1.156(5)  & 1.164(4)  &  1.162(5)  & 1.165(4) \cr
    & $h_+$ & 0.851(12) & 0.838(12) & 0.825(8)  & 0.815(10) \cr
\hline\hline
\end{tabular}
\end{table}

Although the combined fit method helps in many aspects, it requires the handling of a larger data set and 
correspondingly a much larger covariance matrix. 
We use the jackknife method with single elimination to calculate the covariance matrices because the data 
samples show a very small autocorrelation time (less than 1). 
To reduce the time searching for the minimum of $\chi^2$, we take the output of the two-step fit as the 
initial guess for the combined~fit. 

It is worth mentioning a small complication.
When fitting the single ratio $a_i$, we need in advance both the ground-state energy $E_0$ and the 
wave function normalization factor $Z$ to suppress the time dependence of the ratio. 
With the combined fit procedure, $E_0$ and $Z$ are refined through the two-point functions which are part of 
the combined fitting. 
So in the actual analysis, we take the results of $E_0$ and $Z$ from the combined fit and plug them back to 
suppress the time dependence of the three-point function for $a_i$. 
We need to iterate such a process a few times until the fitting results stabilize. 
We find that this iteration converges within two or three steps. 


\section{Chiral-Continuum Extrapolation}
\label{sec:chiral}

Given the light-quark masses in Table~\ref{tab:ensembles}, we extrapolate the results to the physical value
guided by chiral effective theory.
In the case of $h_+(w)$, we follow rooted staggered chiral perturbation theory
(rS$\chi$PT)~\cite{Lee:1999zxa,Aubin:2003mg,Aubin:2003uc,Sharpe:2004is,Aubin:2005aq}.
The specific application to the case of $B\to D^{(*)}$ at zero recoil is provided in
Refs.~\cite{Laiho:2005ue,Bernard:2008dn}.
The continuum $\chi$PT for the semileptonic $B\to D^{(*)}$ form factor at nonzero recoil has been derived
at next-to-leading order (NLO) in Ref.~\cite{Chow:1993hr}, and the generalization to rS$\chi$PT for $B\to D$
is given in the Appendix.
In the case of $h_-(w)$, the leading correction is simply a constant that is inversely proportional to the
charm quark mass.
To describe the simulated data, we also must parametrize the recoil dependence to quadratic order around zero
recoil.
The expansion coefficients are related to the slope and curvature of the form factors.

Thus, we can write the general expression for $h_\pm(w)$ with NLO rS$\chi$PT and 
higher-order analytic terms incorporating lattice-spacing dependence as 
\ben
    h^{\mathrm{lat}}_+(w) &=& 1 -\rho_+^2 (w-1) + k_+ (w-1)^2+\frac{ X_+(\Lambda_\chi)}{m_c^2} +
        \frac{\gDDp^2}{16\pi^2f^2} \mathrm{logs}_{\text{1-loop}}(\Lambda_\chi,w)
    \nonumber \\ &  &  { } + 
        c_{0,+}\, m_x+ c_{1,+}\,(2m_l + m_h) + c_{a,+} a^2, 
    \label{eq:chipt:h+} \\
    h^{\mathrm{lat}}_-(w) & = &  \frac{X_-}{m_c} - \rho_-^2 (w-1) + k_-(w-1)^2  + 
        c_{0,-}\, m_x+ c_{1,-}\,(2m_l + m_h) + c_{a,-} a^2, 
    \label{eq:chipt:h-}
\een
where $-\rho^2_\pm$ and $2k_\pm$ are the slopes and curvatures of the form factors, while $X_\pm$ are
low-energy constants.
In the case of $h_+(w)$, $X_+$ depends on the chiral scale $\Lambda_\chi$ in such a way as to cancel the
$\Lambda_\chi$ dependence of the nonanalytic terms (``chiral logs'').
These terms, denoted here as $\mathrm{logs}_{\text{1-loop}}$, are given by the terms appearing inside the
square brackets in Eqs.~(\ref{eq:xlog_bd}) and~(\ref{eq:xlog_bs}) of the Appendix.
The other form factor $h_-(w)$ has no nonanalyticity at one loop.
Equations~(\ref{eq:chipt:h+}) and~(\ref{eq:chipt:h-}) also include terms depending linearly on the valence
($m_x$) and sea ($m_l$ and $m_s$) quark masses, with coefficients $c_{0(1),\pm}$.
These terms are next-to-next-to-leading order in the chiral expansion and are needed to describe the data
with $m_x$ or $m_l\gtrsim\frac{1}{2}m_s$.
Generic lattice-spacing dependence is described by the terms with coefficients~$c_{a,\pm}$.

We treat the chiral extrapolations of the $B\to D$ and $B_s\to D_s$ data slightly differently. 
For $B\to D$, we analyze only full QCD data points, i.e., $m_x=m_l$. 
Then, since the strange sea quark in all ensembles is tuned within several per cent of its physical mass, 
$m_h\approx m_s$, the dependence of the form factors on the sea and valence quark masses cannot be 
disentangled.
Therefore, we drop the parameter $c_{1,+}$ when fitting the $B\to D$ data. 
For our $B_s\to D_s$ data, on the other hand, the strange valence quark is tuned close to its
physical mass for all the ensembles we analyze, $m_x=m_s$.
As a result, we cannot disentangle the valence quark dependence.
Therefore, we discard the parameter $c_{0,\pm}$ when fitting the $B_s\to D_s$ data, and estimate the
tuning error of $m_s$ \emph{a~posteriori} in Sec.~\ref{subsec:mlms}.
\begin{figure}[bp]
    \centering
    \includegraphics[width=0.48\textwidth, trim=24pt 0 24pt 0]{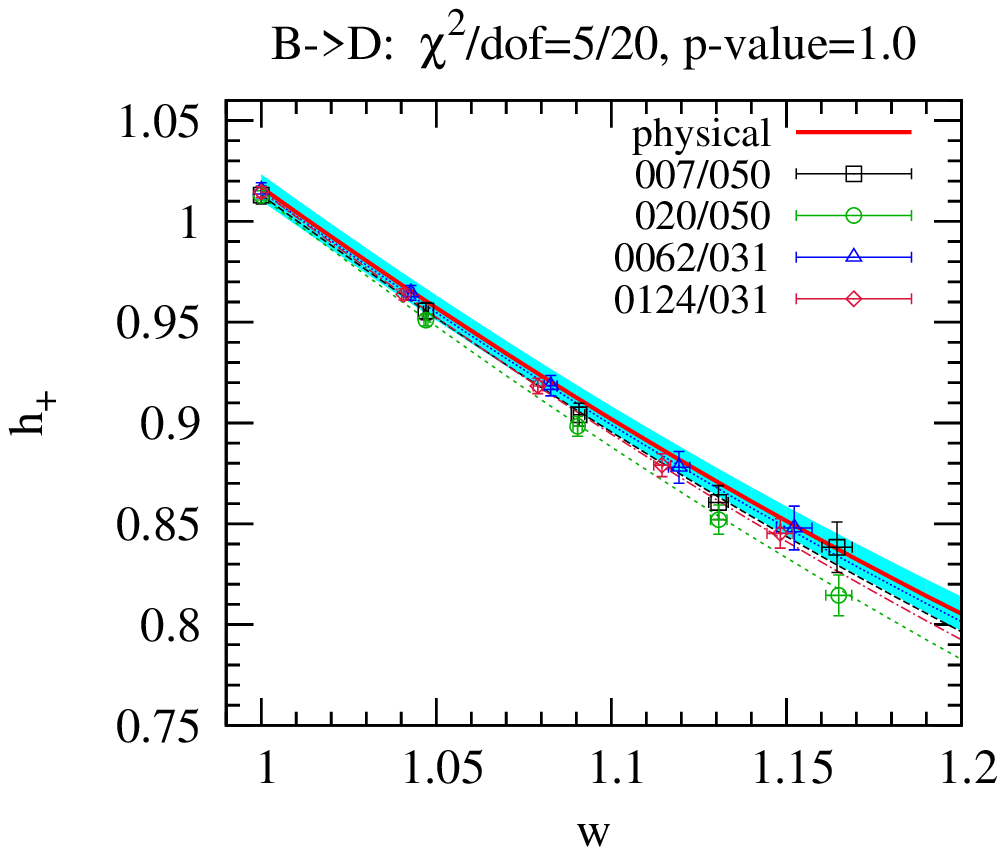}\hfill
    \includegraphics[width=0.48\textwidth, trim=24pt 0 24pt 0]{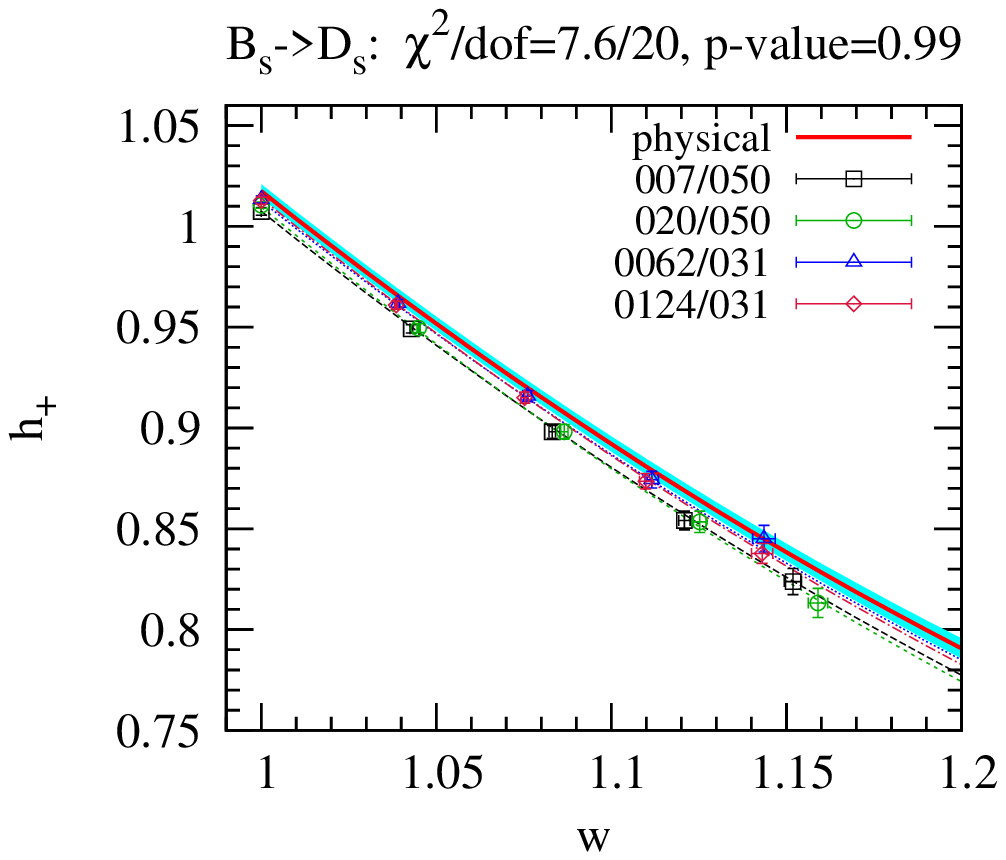}
    \caption{Chiral-continuum extrapolation of $h_+(w)$ for $B\to D$ (left) and $B_s\to D_s$ (right) 
        decays based on the four ensembles listed in Table \ref{tab:ensembles}. 
        The blue bands show only the statistical errors and the red curves are the chiral and continuum 
        limits.}
    \label{fig:chiralhplus}
\end{figure}

The rS$\chi$PT expression for $\mathrm{logs}_{\text{1-loop}}(\Lambda_\chi,w)$ contains several 
low-energy constants used in rS$\chi$PT to describe the masses and decay constant of light pseudoscalar 
mesons.
The values we use for these parameters are taken from Refs.~\cite{Bazavov:2009bb,Bazavov:2010hj} and given 
in Table~\ref{tab:chiralparameters}.
\begin{table}[tp]
    \caption{Input parameters for the chiral extrapolation~\cite{Bazavov:2009bb,Bazavov:2010hj}.}
    \label{tab:chiralparameters}
    \begin{tabular}{ccccc}
    \hline \hline
    rs$\chi$PT & \multicolumn{4}{c}{Ensemble}  \\
    quantity   & 
    C020 & 
    C007 & 
    F0062  &
    F0124  \cr
    \hline 
    $r_1/a$\;    &\; 2.821123 & 2.738591 & 3.857729 & 3.788732  \cr
    $\mu_0$     &\; 6.234000 & 6.234000 & 6.381592 & 6.381592  \cr 
    $r_1^2a^2\Delta_P$\; &\; 0        & 0        & 0        & 0                \cr 
    $r_1^2a^2\Delta_A$\; &\; 0.2052872& 0.2052872& 0.0706188& 0.0706188		  \cr
    $r_1^2a^2\Delta_T$\; &\; 0.3268607& 0.3268607& 0.1153820& 0.1153820        \cr
    $r_1^2a^2\Delta_V$\; &\; 0.4391099& 0.4391099& 0.1523710& 0.1523710        \cr
    $r_1^2a^2\Delta_I$\; &\; 0.5369975& 0.5369975& 0.2062070& 0.2062070        \cr 
    $r_1^2a^2\delta'_V$\;&\; $-0.05$  & $-0.05$  & $-0.03$  & $-0.03$          \cr
    $r_1^2a^2\delta'_A$\;&\; $-0.28$  & $-0.30$  & $-0.15$  & $-0.16$          \cr
    \hline\hline
    \end{tabular}
\end{table}
The $\chi$PT expressions also require the $D_{(s)}$-$D_{(s)}^*$ splitting $\Delta^{(c)}$ and the pion decay 
constant $f_\pi$; we take both from Ref.~\cite{Nakamura:2010zzi}.
To combine data from both lattice spacings, we convert dimensionful quantities to $r_1$ units,
where $r_1$ is the distance defined via the interquark force by 
$r_1^2F(r_1)=1$ \cite{Bernard:2000gd,Sommer:1993ce}. 
We take $r_1/a$ from Refs.~\cite{Bazavov:2009bb,Bazavov:2010hj}.

Unfortunately, the $D^*$-$D$-$\pi$ coupling $\gDDp$ and similar couplings with strange mesons, which appear
in the coefficient $\gDDp^2/16\pi^2f_\pi^2$ of the chiral log terms, are not known with good precision.
We appeal to various estimates of $\gDDp$ available in the literature, including CLEO's measurement of the
$D^*$ width: $\gDDp=0.59(7)$~\cite{Anastassov:2001cw}; quenched lattice QCD:
$\gDDp=0.67(8)(^{+4}_{-6})$~\cite{Abada:2002vj}; a fit to various experimental data, including the $D^*$
width: $\gDDp=0.51$ (no error reported)~\cite{Arnesen:2005ez}; two-flavor lattice QCD in the static limit:
$\gDDp=0.516(51)$~\cite{Ohki:2008py}; and 2+1-flavor lattice QCD in the static limit:
$\gDDp=0.449(51)$~\cite{Detmold:2011bp}.
In this calculation, we include $\gDDp$ as a parameter in the constrained fit with a prior $0.51\pm 0.20$.

In Sec.~\ref{sec:sims}, we compared the two fitting procedures, two-step fit and combined fit, with which we
obtain the single ratios $a_i$, $b_i$, $d_i$ and the double ratio $R_+$.
At each $w$ where we have data, $h_+(w)$ from the two procedures are in good agreement (within $1\sigma$).
We then fit the resulting $h_+(w)$ for the coarse ensembles (C020 and C007) to Eq.~(\ref{eq:chipt:h+})
without the analytic terms and the $a^2$ dependence (NLO).
The results are shown in Table \ref{tab:2step_combined}.
It is apparent that the results from the combined fit procedure are better described by the chiral effective
theory that we employ, giving a $\chi^2/\textrm{d.o.f.} = 0.53$ for $B\to D$ (compared to 1.6 from the
two-step fit).
A similar observation can be found in the case of $B_s\to D_s$.
This indicates that the correlations among the ratios and those among different kinematic points are better
resolved by the combined fit, and hence we follow this procedure for the entire analysis.
\begin{table}[tp]
\caption{\label{tab:2step_combined} Chiral extrapolation of the two-step and combined fit results on the
coarse ensembles.}
\begin{tabular}{c@{\quad}c@{~~}c@{\quad}c@{~~}c}
\hline
\hline
Form	& \multicolumn{2}{c}{Two-step fit} & \multicolumn{2}{c}{Combined Fit} \\	
 factor &   $\chi^2$/d.o.f. & $p$~value    &   $\chi^2$/d.o.f. & $p$~value \\
\hline 
$h_+^{B\to D}(w)$        & 11/7 & $0.14$ & 3.7/7 & 0.81 \cr
$h_+^{B_s\to D_s}(w)$    & 11/7 &  0.13  & 5/7 & 0.68\cr 
\hline
\hline
\end{tabular}
\end{table}
\begin{figure}[bp]
    \centering
    \includegraphics[width=0.48\textwidth, trim=24pt 0 24pt 0]{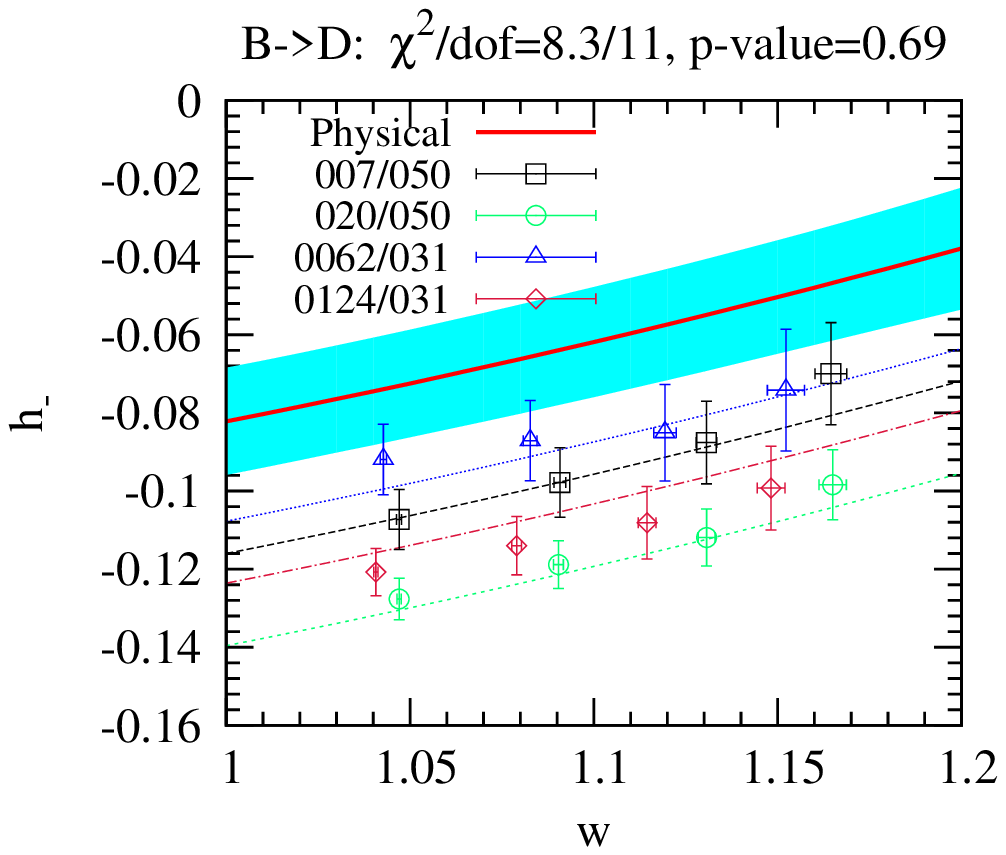}
    \includegraphics[width=0.48\textwidth, trim=24pt 0 24pt 0]{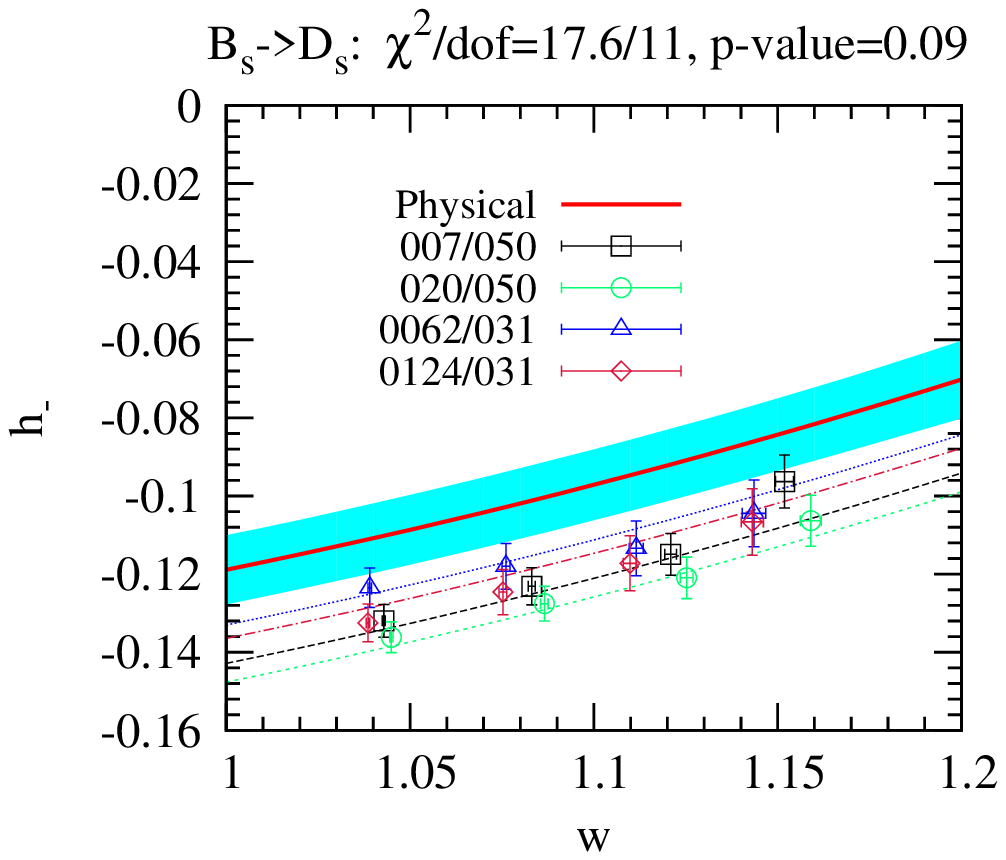}
    \caption{Chiral-continuum extrapolation of $h_-(w)$ for $B\to D$ (left) and $B_s\to D_s$ (right) decays 
        based on the four ensembles listed in Table \ref{tab:ensembles}. 
        The blue bands show only the statistical errors and the red curves are the chiral and continuum 
        limits.}
    \label{fig:chiralhminus}
\end{figure}

The results of the chiral-continuum extrapolation of $h_+(w)$ for $B\to D$ and $B_s\to D_s$ are plotted
in Fig.~\ref{fig:chiralhplus}.
With the large number of configurations we have for the four ensembles, we are able to determine the form
factors $h_+(w)$ with statistical errors at the level of $\sim 0.5\%$ at zero recoil, increasing to 
$\sim 1.5\%$ at $w=1.15$.
The form factor $h_+(w)$ for both of the $B\to D$ and $B_s\to D_s$ decays exhibits a small dependence on the 
light-quark masses and lattice spacings, so the extrapolated physical values are close to the lattice data.
The difference between $h_+^{B\to D}(w)$ and  $h_+^{B_s\to D_s}(w)$ is also small.
The form factor with strange spectator $h_+^{B_s\to D_s}(w)$ shows a steeper slope and larger curvature---%
$\rho^2_+= 1.26(09)$ and $k_+=1.15(9)$---than its $B\to D$ counterpart---$\rho^2_+= 1.14(10)$ and 
$k_+= 0.87(13)$. 

The results of the chiral-continuum extrapolation of $h_-(w)$ for $B\to D$ and $B_s\to D_s$ are plotted in
Fig.~\ref{fig:chiralhminus}.
Here light-quark mass (both sea and spectator) and lattice spacing dependence are visible.
We find an $\approx 0.04$ difference between the two values of $h_-(1)$ at zero recoil.
From Eq.~(\ref{eq:f0}), however, this effect does not cause much difference in~$f_0$.
We therefore anticipate that the $U$-spin symmetry breaking effect is smaller than what was found in
Ref.~\cite{Blasi:1993fi}.
Such an observation is in accord with the recent lattice calculations of $f_+, f_0$ in the $D_{(s)}\to \pi
(K)$ decays \cite{Koponen:2011ev}.
\begin{figure}[bp]
    \centering
    \includegraphics[height=2.75in, trim=24pt 0 48pt 0]{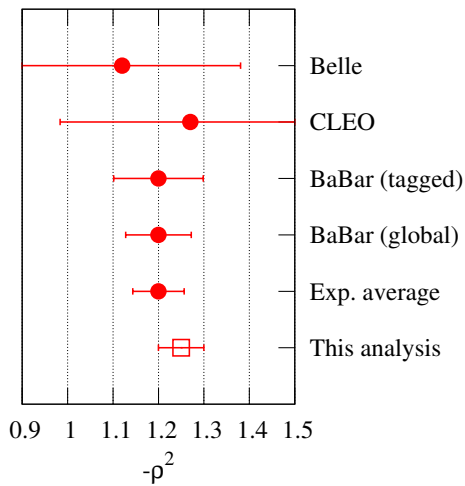}
    \caption{The slope of the form factor $\mathcal{G}(w)$ from the chiral extrapolation is compared with
        various experimental measurements from Belle~\cite{Abe:2001yf}, 
        CLEO~\cite{Bartelt:1998dq}, BaBar (tagged)~\cite{babar:2008ii}, and 
        BaBar (global)~\cite{Aubert:2008yv} respectively.}
    \label{fig:rho_compare}
\end{figure}

With the results from the chiral-continuum extrapolation in hand, we now convert $h_\pm(w)$ into $f_+(q^2)$
and $f_0(q^2)$ with Eqs.~(\ref{eq:f+}) and (\ref{eq:f0}) and the physical $B$ and $D$
masses~\cite{Nakamura:2010zzi}.
To gain confidence in our procedures, let us compare the resulting $f_+$ with experimental measurements.
The differential decay rate of $B\to D$ is given by
\be
    \frac{d\Gamma (\bar{B}^0 \to D\ell\bar{\nu})}{dw} =
        \frac{G_F^2}{48\pi^3}M_D^3(M_B+M_D)^2(w^2-1)^{3/2}|V_{cb}|^2|\mathcal{G}(w)|^2,
\ee 
where $G_F$ is the Fermi constant, and it is conventional to introduce
\be
    \mathcal{G}(w) = \frac{2\sqrt{r}}{1+r} f_+(w).
    \label{eq:G}
\ee
Experiments report the zero-recoil form factor $|V_{cb}|\mathcal{G}(1)$ and the relative form factor slope
$\rho^2\equiv-\mathcal{G}'(1)/\mathcal{G}(1)$~\cite{luth:2011}.
From our extrapolated data, we find $\mathcal{G}(1)=1.058(9)_{\textrm{stat.}}$, which is consistent with the
previous unquenched lattice-QCD result
$1.074(18)_{\textrm{stat.}}(16)_{\textrm{syst.}}$~\cite{Okamoto:2004xg}.
The measured slope is related to parameters of our chiral extrapolation via
\be
    \rho^2 = \frac{1}{\mathcal{G}(1)}\; \left[ \rho^2_+ + \frac{r-1}{r+1}\;\rho_-^2  + 
        \rho^2_{\text{logs}} \right] ,
\ee
where $\rho^2_{\textrm{logs}}$ is the slope of the NLO logarithm at zero recoil. 
In Fig.~\ref{fig:rho_compare}, we compare the slope of the $B\to D$ form factor at $w=1$ with experiment. 
We find $\rho^2=1.25(5)_{\textrm{stat.}}$, where the error is obtained from 300 bootstrap samples.
This value is in good agreement with the experimental results from
Belle, CLEO and BaBar \cite{luth:2011} and the average of these from the Heavy Flavor Averaging Group,
$\rho^2=1.18(6)$~\cite{Asner:2010qj}.

Note that a determination of $\mathcal{G}(w)$ with full error budget is beyond the scope of this paper.
A~comprehensive effort to do so is in progress~\cite{Qiu:2011ur}.
Here we are merely satisfied to see that the main ingredients of our analysis are compatible with the 
available experimental data.


\section{\boldmath $z$ Parametrization}
\label{sec:z}

To minimize discretization effects, the final $D$ meson momentum $\bm{p}$ should not be taken too large, so
the calculations are restricted to small recoil, $w<1.17$.
However, the form-factor ratio that we are trying to compute ultimately needs to be evaluated near maximum
recoil, $w\sim 1.6$, which appears to require a considerable extrapolation.
Fortunately, the extrapolation can be guided by the model-independent
$z$~parametrization~\cite{deRafael:1993ib,Boyd:1995sq,Boyd:1997kz}.
As shown, for example, in Ref.~\cite{Bailey:2008wp}, this strategy is effective for extrapolating lattice-QCD
data.

One introduces the variable
\be
    z(w) = \frac{\sqrt{1+w}-\sqrt{2}}{\sqrt{1+w}+\sqrt{2}},
\ee
which maps the physical domain of the form factors into the unit disk. 
The form factors can be expressed as
\begin{equation}
    f_{i}(z) = \frac{1}{P(z) \phi(z)} \sum_{n=0}^{\infty} a_n z^n
    \label{eq:PphiF}
\end{equation}
where $P(z)$ and $\phi(z)$ are called, respectively, the Blaschke factor and the outer function.
The range of $w$ for $B\to D$ is $1\le w \le 1.589$, which corresponds to $0 \le z \le 0.0644$.
With the choice of outer functions given below, unitarity sets a bound on the expansion coefficients,
\begin{equation}
    \sum_n |a_n|^2 \le 1. \label{eq:unitarity}
\end{equation}
Because of this constraint and the restricted range of $z$, the expansion in Eq.~(\ref{eq:PphiF}) converges, 
and one can parametrize the form factors with only a few terms. 
In this analysis, we truncate the expansion at the $z^2$ term, which is enough for our data.

Although we are primarily interested in $f_0$, we apply the $z$~expansion to $f_+$ and $f_0$ simultaneously,
incorporating the kinematic constraint $f_0(0)=f_+(0)$.
We take the outer functions to be
\ben
    \phi_0(z) & = & \Phi_0 \;(1+z)\; (1-z)^{3/2} \left[(1+r)(1-z) + 2\sqrt{r}(1+z)\right]^{-4},
    \label{eq:outer0} \\
    \phi_+(z) & = & \Phi_+\; (1+z)^2 (1-z)^{1/2} \left[(1+r)(1-z) + 2\sqrt{r}(1+z)\right]^{-5}.
    \label{eq:outer+}
\een 
where we choose the constants $\Phi_0 = 0.5299$ and $\Phi_+ = 1.1213$ such that it matches the same
normalization as in Ref.~\cite{Boyd:1995sq}.
These exact values can be combined with the fit results, given below, to reconstitute the form factors.

\begin{figure}[bp]
    \centering
    \includegraphics[width=0.48\textwidth, trim=24pt 0 24pt 0]{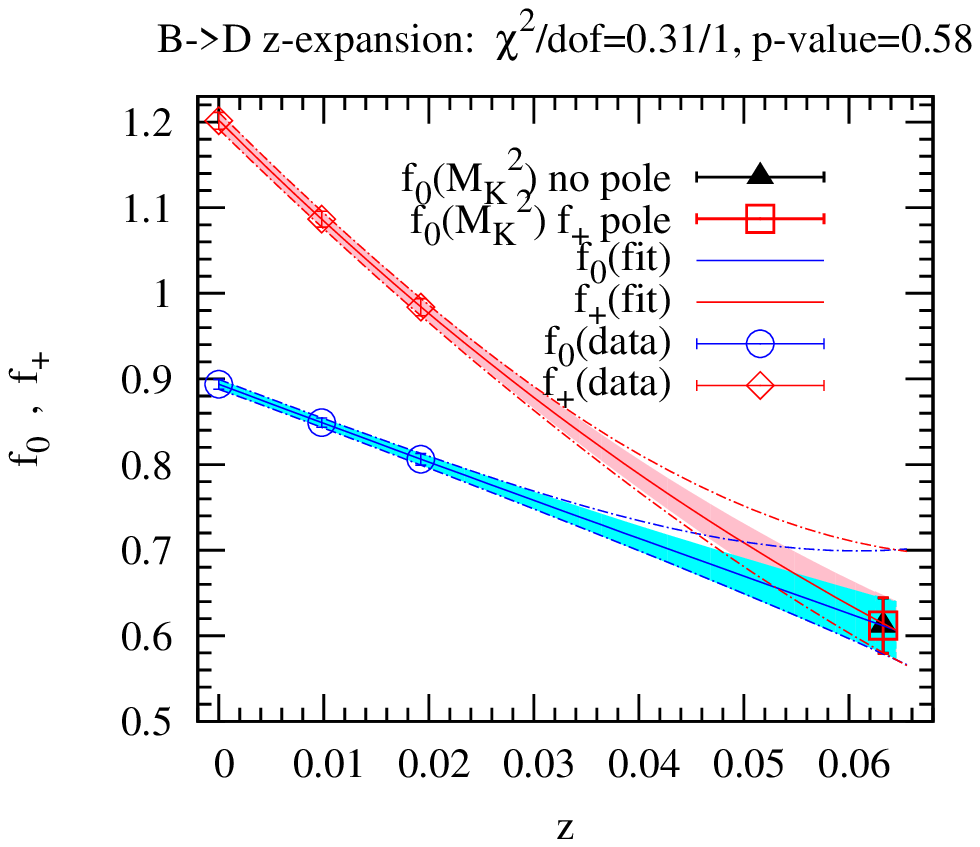}\hfill
    \includegraphics[width=0.48\textwidth, trim=24pt 0 24pt 0]{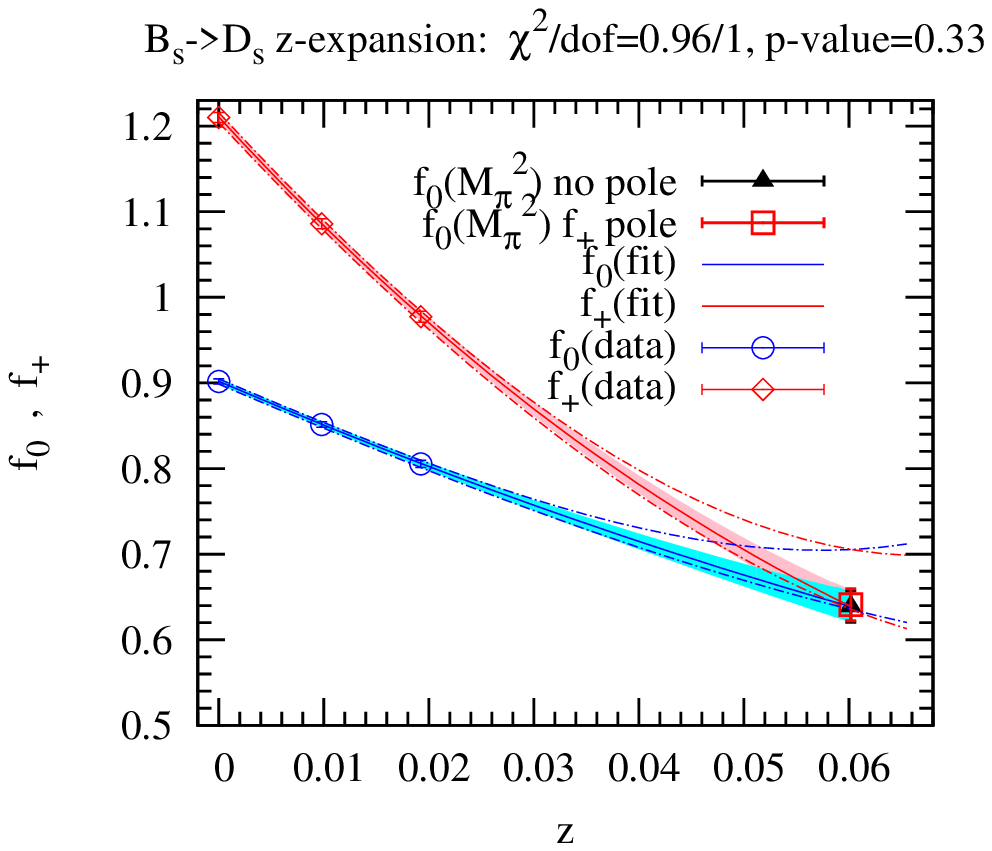}
    \caption{$z$~expansion of form factors to the maximum recoil point for both $B\to D$ and $B_s\to D_s$.
        The diamonds and circles for $z\le0.02$ are the synthetic data derived from the chiral-continuum 
        extrapolation, and the solid curves and error band are the results of the $z$~expansion.
        The dashed curves show how the chiral-continuum extrapolation extends into the region where the 
        extrapolation may not be trustworthy. 
        The points near $z=0.06$ correspond to the desired $q^2=M_\pi^2$ and $M_K^2$;
        squares (triangles) correspond to $z$ fits with (without) the $B_c^*$ pole in the Blaschke factor.}
    \label{fig:zfit}
\end{figure}
\begin{figure}[bp]
    \centering
    \vspace*{-3pt}
    \includegraphics[width=4in]{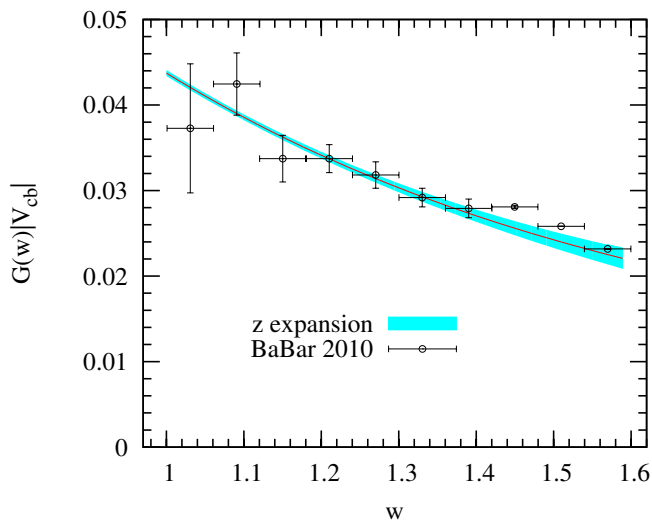}
    \vspace*{-3pt}
    \caption[tab:gw_compare]{Comparison of the form factor shape of $\mathcal{G}(w)|V_{cb}|$ with BaBar's 
        measurements\vphantom{\cite{Gregory:2009hq}}~\cite{Aubert:2009ac}.}
    \label{fig:gw_compare}
\end{figure}
The purpose of the Blaschke factor is to remove poles outside the physical region, i.e., at~$w<1$.
In general,
\be
    P(z) =  \prod_{p=1}^{N_P} \frac{z(w) - z(w_p)}{1-z(w)\,z(w_p)},
    \label{eq:Blaschke}
\ee
where $z(w_p)$, with $w_p=[1+r^2-M_p^2/M^2_{B_{(s)}}]/2r$, marks the pole position, and the product can run 
over states with $M_p<M_{B_{(s)}}+M_{D_{(s)}}$. 
In the case at hand, these poles appear at the masses of $J^P=1^-$ vector ($J^P=0^+$ scalar)
$B_c$ mesons for $f_+$ ($f_0$). 

We need to put the results of the chiral-continuum extrapolation in the $z$-parametrized form of
Eq.~(\ref{eq:PphiF}).
We do so by generating synthetic data from the chiral-continuum extrapolated curves and fitting for the
corresponding parameters $a_i$ in Eq.~(\ref{eq:PphiF}).
Our chiral-continuum fit has eleven free parameters, including ones to describe the $a^2$ dependence
in~$h_\pm(w)$.
Such terms vanish in the continuum limit, leaving only nine physical parameters.
Furthermore, due to the small contribution from the terms of $\gDDp$ and $c_{0(1),+}$ and due to the
correlation between $h_\pm(w)$, we end up with effectively six free modes in the synthetic data.
Thus, we generate the synthetic data at three evenly-spaced values $w\in\{1.0, 1.08,1.16\}$ in the
region where we have data.
In Sec.~\ref{sec:z-syst}, we show that the final form-factor ratio is not sensitive to the details of these
synthetic data.
We start with trivial Blaschke factors (no poles) and fit the $B\to D$ and $B_s\to D_s$ form factors up
to the $z^2$ term in Eq.~(\ref{eq:PphiF}).
The results of these fits are shown in Fig.~\ref{fig:zfit}.
To examine the effect of poles on the shape of the form factor $f_+$, we also try a fit with a one-pole
Blaschke factor at mass $M_p=M_{B_c^*}=6.330$~GeV, where this lowest $B_c^*$ is a prediction of lattice
QCD~\cite{Gregory:2009hq}.
This Blaschke factor affects the extrapolated results with a deviation of about 0.3\%.
The dashed lines in Fig.~\ref{fig:zfit} show the extrapolation based solely on the chiral fit, showing that
the $z$ expansion plays an important role in controlling the total error.

We perform the $z$-expansion fit without constraints on $a_0$, $a_1$, and $a_2$, and setting the rest to zero.
In Sec.~\ref{sec:z-syst}, we discuss fits with more parameters, constrained then by Eq.~(\ref{eq:unitarity}).
We constrain the fit with the relation $f_0(0) = f_+(0)$,
by demanding $|f_0(0)- f_+(0)| < \delta$ where $\delta$ can be chosen arbitrarily small.
Once $\delta$ is small enough, its actual value has no effect on the fit. 

To check the form factor shape obtained from the $z$~expansion, we can compare the $B\to D$ decay with the
latest published measurement from BaBar~\cite{Aubert:2009ac}.
The comparison is shown in Fig.~\ref{fig:gw_compare}, using $|V_{cb}|=41.4\times10^{-3}$, as determined from 
$B\to D\ell\nu$ at nonzero recoil~\cite{Aubert:2009ac,deDivitiis:2007ui}.
As one can see, the shape of our form factor prefers a larger value of $|V_{cb}|$ and agrees well with 
experiment over the full kinematic range.
As above, we note that this comparison is made without a full treatment of the systematic errors on 
$\mathcal{G}(w)$.
A thorough treatment with full error analysis, aimed at determining $|V_{cb}|$, will be covered elsewhere;
see Ref.~\cite{Qiu:2011ur} for a progress report. 

For completeness, we give the results of the $z$ fit in Table~\ref{tab:anz}.
\begin{table}[tp]
    \caption[tab:anz]{Best-fit values $a_n$ and correlation matrix $\rho_{mn}$ of the simultaneous 3-term 
        $z$~expansion of $f_+$ and $f_0$, with statistical (post extrapolation) errors only.
        Top:~$B\to D$; bottom:~$B_s\to D_s$. 
        Note that the fit parameters are correlated between the $B\to D$ and $B_s\to D_s$ processes.}
    \label{tab:anz}
    \begin{tabular}{ccccccc}
        \hline\hline
            $B\to D$:\quad & 0.0126(1)& $-0.106(4)$ & 0.32(9) &  0.01130(7) & $-0.061(4)$ & 0.03(10)  \\
        \hline $\rho$   
            & \multicolumn{1}{c}{$a_0^{(+)}$} & \multicolumn{1}{c}{$a_1^{(+)}$} & \multicolumn{1}{c}{$a_2^{(+)}$}
            & \multicolumn{1}{c}{$a_0^{(0)}$} & \multicolumn{1}{c}{$a_1^{(0)}$} & \multicolumn{1}{c}{$a_2^{(0)}$} \\
            $a_0^{(+)}$ & 1.000 & $-0.273$ & $-0.012$ & \hs0.664 & $-0.061$ & \hs$ -0.014$ \\
            $a_1^{(+)}$ &  & \hs1.000 & $-0.293$ & $-0.306$ & \hs$0.917$ & $-0.164$ \\
            $a_2^{(+)}$ &  &  & \hs1.000 & $0.045$ & $-0.311$ & \hs$0.976$ \\
            $a_0^{(0)}$ &  &  &  & \hs1.000 & $-0.299$ & $0.009$ \\
            $a_1^{(0)}$ &  &  &  &  & \hs1.000 & $-0.231$ \\
            $a_2^{(0)}$ &  &  &  &  &  & \hs1.000 \\
\hline
            $B_s\to D_s$: & 0.01191(6)  & $-0.111(2)$ & 0.47(5) & 0.01081(4) & $-0.066(2)$ & 0.18(6)  \\
        \hline $\rho$ 
            & \multicolumn{1}{c}{$a_0^{(+)}$} & \multicolumn{1}{c}{$a_1^{(+)}$} & \multicolumn{1}{c}{$a_2^{(+)}$}
            & \multicolumn{1}{c}{$a_0^{(0)}$} & \multicolumn{1}{c}{$a_1^{(0)}$} & \multicolumn{1}{c}{$a_2^{(0)}$} \\
            $a_0^{(+)}$ & 1.000 & $-0.055$ & $-0.002$ & \hs0.593 & $0.254$ & $\hs0.014$ \\
            $a_1^{(+)}$ &  & \hs1.000 & $-0.318$ & \hs$-0.067$ & \hs0.867  & $-0.180$ \\
            $a_2^{(+)}$ &  &  & \hs1.000 & \hs$-0.038$ & $-0.307$ & \hs0.974 \\
            $a_0^{(0)}$ &  &  &  & \hs1.000 & $-0.050$ & \hs$-0.054$  \\
            $a_1^{(0)}$ &  &  &  &  & \hs1.000 & $-0.233$ \\
            $a_2^{(0)}$ &  &  &  &  &  & \hs1.000 \\
        \hline\hline
    \end{tabular}
\end{table}
The correlation matrix does \emph{not} include full systematics, but with this information, the reader can 
reproduce the curves and error bands in Figs.~\ref{fig:zfit} and~\ref{fig:gw_compare}.
Note, however, that the parameters of the nonstrange and strange form factors are also correlated.


\section{Systematic Errors}
\label{sec:sys}

We now discuss the systematic errors in our analysis.
Owing to the similarity between the $B\to D$ and $B_s\to D_s$ processes, the systematic errors in the ratio
of the form factors largely cancel, by design.
To assess the systematic uncertainties, we have repeated the chiral-continuum and $z$ extrapolations with
different choices.
The values of $f^{(s)}_0(M^2_\pi)$, $f^{(d)}_0(M^2_K)$, and $f^{(s)}_0(M^2_\pi)/f^{(d)}_0(M^2_K)$ resulting
from these variations are listed in Table~\ref{tab:f0_all}.
We summarize the final error budget in Table~\ref{tab:error_budget}.
As our standard analysis, we use Eqs.~(\ref{eq:chipt:h+}) and~(\ref{eq:chipt:h-}) for the chiral-continuum 
fit, dropping $c_{1,\pm}$ for $B\to D$ and $c_{0,\pm}$ for $B_s \to D_s$.
We fit the coupling $\gDDp$ using a constrained fit~\cite{Lepage:2001ym} with the prior $0.51(20)$.
We take the synthetic data points at $w=1.0,1.08$ and $1.16$ and include the trivial Blaschke factor (no
poles).
We use $r_1=0.3117$~fm to convert the necessary physical inputs (like $f_\pi$) to $r_1$~units.

\subsection{Chiral extrapolation}

Our chiral extrapolation is based on the rS$\chi$PT formalism shown in
Eqs.~(\ref{eq:chipt:h+}) and~(\ref{eq:chipt:h-}) and the Appendix.
The systematic errors arising here can be divided into two categories: the error associated with the 
one-loop contribution itself and the error associated with the partial inclusion of NNLO analytic terms.
Throughout our analysis, we keep the slope $\rho^2$ and curvature $k$, because they determine the $w$ 
dependence of the form factors.

An uncertainty in the contribution from the NLO logarithm stems from the uncertainty of the $D^*$-$D$-$\pi$
coupling $\gDDp$.
Although the NLO logarithms to $h_+(w)$ make a small contribution when evaluated with the quark masses for
which we have data, of order $10^{-3}$ at our lightest simulated quark mass, the logarithm affects the
form factor shape of $h_+(w)$ through its $w$~dependence.
Thus, the uncertainty in $\gDDp$ becomes more important as we extend our calculations to large recoil.
We include $\gDDp$ in the chiral-continuum fit with the prior $0.51(20)$, which describes the data well, so
we do not assign an additional error due to the uncertainty of~$\gDDp$.

\begin{table}[tp]
    \caption{Values of the form factors $f_0^{(s,d)}$ and their ratio for several variants of the fitting 
        procedure. 
        The second panel shows the results for different choices of chiral extrapolation fit function.  
        Note that for the $B \to D$ form factor we use the fit functions labeled ``val" while for the 
        $B_s\to D_s$ form factor we use the fit functions labeled ``sea" in 
        Eqs.~(\ref{eq:NLO+})--(\ref{eq:NNLO:all}).  
        The third panel shows the results for different choices of the $z$~expansion.
        The final panel shows the results for different values of parametric inputs: the lattice scale and 
        light- and strange-quark masses.}
    \label{tab:f0_all}
    \begin{tabular}{l@{\quad}c@{\quad}c@{\quad}c}
    \hline\hline
    Variations      & 
    $f^{(s)}_0(M^2_\pi)$ & $f^{(d)}_0(M^2_K)$ & $f^{(s)}_0(M^2_\pi)/f^{(d)}_0(M^2_K)$ \cr
    \hline 
    Standard (pNNLO$^{\pm}_{w,a,{\rm sea/val}}$)         &   0.639(19)   & 0.612(32)   & 1.046(44) \cr
    \hline
    NLO$_w^\pm$      &   0.636(17)   & 0.618(30)   & 1.031(42) \cr
    pNNLO$_{w,\text{sea/val}}^\pm$ &  0.633(18)  & 0.616(31)   & 1.030(43) \cr
    $\text{NLO}_w^+\oplus\text{pNNLO}_{w,\text{sea/val}}^-$ &  0.623(18)  & 0.594(31)   & 1.051(45) \cr
    \hline
    With $B_c^*$ pole in $f_+$ & 0.641(19) & 0.612(33)  & 1.049(45) \cr 
    $w$-Range[1,1.08]     &   0.623(17)   & 0.600(31)   & 1.042(42) \cr
    $w$-Range[1,1.12]     &   0.631(18)   & 0.606(31)   & 1.044(43) \cr
    $w$-Range[1,1.20]     &   0.646(19)   & 0.618(33)   & 1.048(45) \cr
    $w$-Range[1,1.24]     &   0.653(19)   & 0.623(34)   & 1.049(46) \cr
    Truncated at $z^3$     &   0.632(22)  & 0.607(36)  & 1.042(45) \cr
    Truncated at $z^4$     &   0.632(22)  & 0.608(36)  &  1.043(46) \cr
    \hline
    $r_1=0.321$~fm       &   0.638(18)   & 0.611(32)    & 1.047(44) \cr
    $m_s$ 1$\sigma$ shift   &   0.638(18)   & 0.611(32)      & 1.046(44)       \cr
    $m_l$ 1$\sigma$ shift   &   0.639(18)   & 0.612(32)      & 1.046(44)       \cr
    \hline\hline
    \end{tabular}
\end{table}

We now look at variations from fitting with and without the NNLO analytic terms.
For ease of discussion, let us break the chiral fitting scheme into the following different pieces:
\ben
    \text{NLO}_w^{+} & = &  1 - \rho_+^2 (w-1) + k_+ (w-1)^2+\frac{ X_+}{m_c^2} + 
        \frac{\gDDp^2}{16\pi^2f^2}\mathrm{logs}_{\text{1-loop}}(\Lambda_\chi,w),
    \label{eq:NLO+} \\
    \text{NLO}_w^- &=& 1 - \rho_-^2 (w-1) + k_- (w-1)^2 + \frac{X_-}{m_c},
    \label{eq:NLO-} \\
    \text{pNNLO}^\pm_{w,\text{val}} & = & \text{NLO}_w^\pm + c_{0,\pm} m_x,
    \label{eq:NNLO:val}\\
    \text{pNNLO}^\pm_{w,\text{sea}} & = & \text{NLO}_w^\pm + c_{1,\pm} (2 m_l + m_s),
    \label{eq:NNLO:sea}\\
    \text{pNNLO}^\pm_{w,a,\text{sea/val}} & = & \text{pNNLO}^\pm_{w,\text{sea/val}} + c_{a,\pm}a^2 ,
    \label{eq:NNLO:all}
\een
where ``pNNLO'' stands for partial NNLO, because we include only analytic terms.
We are not aware of any full NNLO calculations with nonzero final $D$-meson momentum. 
\begin{table}[tp]
   \caption{The error budget of the form-factor ratio discussed in text.
       The first row gives the statistical error after the chiral-continuum extrapolation.
       As explained in the text, variations in the chiral functional form make insignificant changes, so we
       quote no extra error for these variations.
       An addition discretization error for heavy-quark effects is in the last row.}
   \label{tab:error_budget}
   \begin{tabular}{lc}
   \hline\hline
   Source of error &
   $\delta(f_0^{(s)}/f_0^{(d)})$
   \cr
   \hline
   Statistics $\oplus$ chiral-continuum~~ & 4.2\% \cr
   $z$ expansion                  & 0.6\% \cr
   Scale $r_1$                    & 0.1\% \cr
   Mistuned $m_s$                 & 0.1\% \cr
   Mistuned $m_l$                 & 0.1\% \cr
   Heavy-quark ($\kappa$) tuning  & 0.6\% \cr
   Heavy-quark discretization     & 1.0\% \cr
   \hline\hline
   \end{tabular}
\end{table}

As mentioned in Sec.~\ref{sec:chiral}, the form factor $h_+(w)$ shows a weak dependence on the light sea and
valence quark masses, and the data are already well-described by NLO$_w^+$, with $\chi^2/\mbox{d.o.f.}=0.68$
and $0.83$, respectively, for $h^{B\to D}_+(w)$ and $h^{B_s\to D_s}_+(w)$.
Adding the pNNLO$^+$ terms, the $h_+$ fits remain good.
As seen in Figs.~\ref{fig:chiralhplus} and~\ref{fig:chiralhminus}, the data for $h_-(w)$ exhibit a more
significant dependence on the lattice spacing and on the sea- and spectator-quark masses.
Unsurprisingly, the NLO$_{w}^-$ fit of $h_-(w)$ leads to large $\chi^2/\text{d.o.f.}=1.7$ and $1.9$,
respectively, for $h^{B\to D}_-$ and $h^{B_s\to D_s}_-$.
The NLO $\chi$PT correction to $h_-(w)$, denoted $X_-/m_c$, does not depend on quark mass or lattice spacing,
so the observed dependence in the data must be described by pNNLO terms.
Moving through the pNNLO functional forms in Eqs.~(\ref{eq:NNLO:val})--(\ref{eq:NNLO:all}), we find that the
pNNLO$^-_{w,\text{val}}$ fit of $h^{B\to D}_-(w)$ improves nicely, $\chi^2/\text{d.o.f.}=0.93$, but
pNNLO$^-_{w,\text{sea}}$ fit of $h^{B_s\to D_s}_-(w)$ less so, $\chi^2/\text{d.o.f.}=1.8$.
These can be contrasted with the standard fits, pNNLO$^\pm_{w,a,\text{val}}$ for $B\to D$ and
pNNLO$^\pm_{w,a,\text{sea}}$ for $B_s\to D_s$, with $h_-$ $\chi^2/\text{d.o.f.}=0.75$ and $1.6$, respectively.
As seen in Table~\ref{tab:f0_all}, these less good fits all lie well within the extrapolated statistical
error of the standard fits.
We therefore treat these alternatives as cross checks and do not add an additional error here.

\subsection{\boldmath $z$ expansion}
\label{sec:z-syst}

Although the $z$ expansion provides a model-independent parametrization of the form factors $f_0$ and $f_+$,
the final results may depend on three kinds of choices made within this framework.
First, the expansion coefficients may depend on the number and range of synthetic data points.
Second, the shape of the form factor may be affected by the number of poles in the Blaschke factor,
particularly for~$f_+$.
Last, the shape may be affected by the truncation of the series in $z$ if one does not include enough terms.

\begin{figure}[bp]
    \centering \vspace*{-0.6em}
    \includegraphics[width=4in]{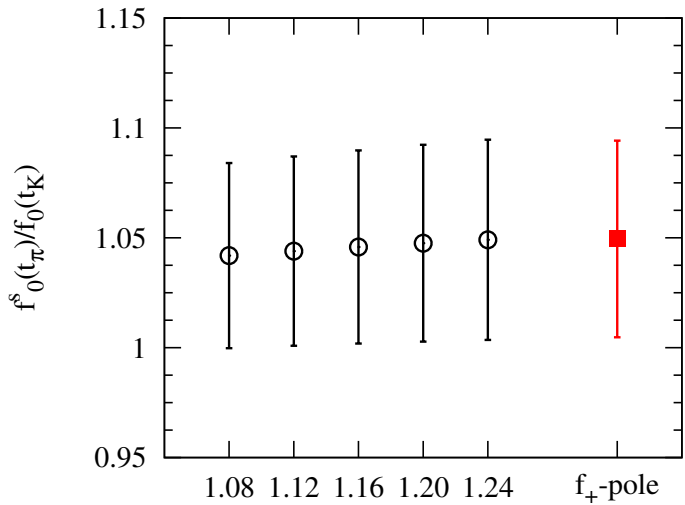}
    \caption{Systematic error due to the $z$~expansion. 
    The open circles show the effect of varying the synthetic data used, and the filled square shows the 
    effect of adding a pole to the Blaschke factor.}
    \label{fig:zfit_systematic}
\end{figure}
To estimate the uncertainty from the synthetic data, we vary the $w$~range over which they are generated.
We repeat the $z$ expansion with $w$ in the intervals $[1,1.08]$, $[1,1.12]$, $[1,1.20]$, $[1,1.24]$,
choosing three evenly-spaced points for both form factors.
We also try fits with more synthetic data than underlying parameters, in which case some of the information
is spurious, leading to tiny eigenvalues in the synthetic-data covariance matrix.
We remove the corresponding mode(s) with singular-value decomposition.
Although the form factors $f_0^{(d,s)}(0)$ each vary with these alternative choices by about $1\sigma$, the
ratio $f^{(s)}_0(M^2_\pi)/f^{(d)}_0(M^2_K)$ is negligibly affected; cf.\ Fig.~\ref{fig:zfit_systematic}.
We take the maximum deviation from $1.046$, which is $0.004$, as the systematic error.

To estimate the uncertainty from the poles in the unphysical region, we repeat the $z$~expansion fit by 
including a $B_c^*$ pole in the Blaschke factor for~$f_+$. 
We take $M_{B_c^*}=6.330$~GeV from a lattice-QCD calculation on the MILC ensembles~\cite{Gregory:2009hq}. 
Recall that $f_+$ influences $f_0$ near maximum recoil via the kinematic constraint $f_0(0)=f_+(0)$.
We find that the effect is rather small, as shown in Fig.~\ref{fig:zfit_systematic}, leading
to a difference of only $0.3\%$ in the ratio of $f_0$. 

To estimate the uncertainty from the truncation of higher order terms in the $z$ expansion, we perform the
$z$~expansion fit by including $z^3$ and $z^4$ terms and incorporating the unitarity constraints on the
coefficients, {\it i.e.}, $a_i^2<1$.
As can be seen in Table \ref{tab:f0_all}, both the form factors and their ratios stabilize when higher order
term $z^3$ (or further $z^4$) is included.
This results in a $0.3\%$ difference in the form-factor ratio.

The total systematic error including all these effects added in quadrature is $0.6\%$.

\subsection{Scale \boldmath $r_1$ dependence}

As discussed in Sec.~\ref{sec:chiral}, we convert our data to $r_1$ units with $r_1/a$ from 
Ref.~\cite{Bazavov:2009bb}.
To convert to physical units, we must choose a value of $r_1$ in physical units.
Our choice is based on MILC's analysis of $f_\pi$, which leads to a mass-independent value 
$r_1=0.3117(6)({}^{+12}_{-31})$~fm \cite{Bazavov:2009bb}. 
To estimate the error, we also consider an early value from HPQCD based on the $2S$-$1S$ splitting of 
the $\Upsilon$ resonances, $r_1=0.321(5)$~fm \cite{Gray:2005ur},
and repeat the whole analysis with this value.
We find a negligible shift, $\sim 0.1\%$, in the form-factor ratio, because of
the cancellation between $B_s\to D_s$ and $B\to D$.

\subsection{Light- and strange-quark mass dependence}
\label{subsec:mlms}

The physical light and strange quark masses are determined from the analysis of light pseudoscalar meson 
masses and decay constants~\cite{Aubin:2004fs,Bazavov:2009bb}.
To estimate the error, we repeat the chiral-continuum extrapolation varying the masses by $\pm1\sigma$.
We do so twice, once varying the physical light quark mass by 3.1\% for $B\to D$; and again varying the 
physical strange-quark mass by 3.4\% for $B_s\to D_s$.
In both cases, we find a shift on $f^{(s)}_0(M^2_\pi)/f^{(d)}_0(M^2_K)$ less than $0.1\%$, 
which is much smaller than other errors in this analysis.

\subsection{Heavy-quark mass dependence}

We have not generated data for a wide-enough range of $\kappa_b$ and $\kappa_c$ to determine directly the 
heavy-quark mass dependence of the $B_s\to D_s$ form factors.
Since our main focus is a $U$-spin breaking ratio, we rely instead on other such ratios computed on the same 
ensembles, in particular the form factor for $B\to D^*$ at zero recoil 
$h_{A_1}(1)$~\cite{Bernard:2008dn,Bailey:2010gb} and the ratio of leptonic decay constants 
$f_{B_s}/f_{B^+}$~\cite{Bazavov:2011aa}.

In the case of $h_{A_1}(1)$, which is very similar to $h_+(1)$, we find the $\kappa$-tuning error to be 
0.56\% of $h_{A_1}(1)$ and 
4.8\%  of $1-h_{A_1}(1)$~\cite{Bailey:2010gb}.
In the case of $\xi_f=f_{B_s}/f_{B^+}$, we find the $\kappa$-tuning error to be 
0.41\% of $\xi_f$ and 
2.2\%  of $\xi_f-1$~\cite{Bazavov:2011aa}.
The first of these four estimates yields the largest absolute error on $f_0^{(s)}/f_0^{(d)}$, namely~$0.6\%$.
This error estimate is still much smaller than the overall error in this analysis.

\subsection{Heavy-quark mass discretization and matching}

We also use our work on $h_{A_1}(1)$~\cite{Bernard:2008dn,Bailey:2010gb} to guide and estimate heavy-quark
discretization errors, both power-law and radiative effects.
For $h_{A_1}(1)$, we find a 1.0\% error from discretization effects, and a 0.3\% error from matching.
Since the present calculation matches only at tree level, the corresponding errors here are order
$\alpha_s$ instead of $\alpha_s^2$.
For a $U$-spin-breaking ratio such as ours, the discretization error is further suppressed by
$(m_s-m_d)/\Lambda_{\rm QCD}$.
Since $(m_s-m_d)/(\alpha_s\Lambda_{\rm QCD})\sim\frac{1}{2}$, there is not much change.
From the structure of Eq.~(\ref{eq:f0}), the matching error stemming from $h_+$ is
$\alpha_s(m_s^2-m_d^2)a^2$, which is negligible, but the matching error from $h_-$ leads to an error on $f_0$
of order $\alpha_s(h_-/h_+)(m_s-m_d)/\Lambda_{\rm QCD}\approx0.5\%$.
Taking a 1\% error for these effects seems reasonable yet does not influence the total error budget much.

\subsection{ Finite-volume effects}

The finite-volume correction to the function defined in Eq.~(\ref{eq:F+}) in the NLO formula at zero recoil
is given in Ref.~\cite{Laiho:2005ue}.
Such correction was found to be very small in the $B\to D^*$ form factor \cite{Bernard:2008dn}.
One should expect similar conclusion in the case of $B_{(s)}\to D_{(s)}$.
Indeed, we find that the largest effect appears at the physical light quark masses and the magnitude of the
correction is $\sim1.0 \times 10^{-4}$ which can be safely ignored.
Although the formula at nonzero recoil is not yet available, we suspect such a correction will cause any
sizable effect at small recoil, considering the fact that the correction at zero recoil is two orders of
magnitude smaller than other systematic errors.
So we do not quote any systematic error from the finite-volume effects.

\subsection{Summary}

Let us now summarize our results.
Table~\ref{tab:f0_all} lists the values of $f^{(s)}_0(M^2_\pi)$, $f^{(d)}_0(M^2_K)$ and their ratio
$f^{(s)}_0(M^2_\pi)/f^{(d)}_0(M^2_K)$ under the variations in the analysis explained above.
The resulting error budget is given in Table~\ref{tab:error_budget}, based on which, we arrive at our final
result given in Eq.~(\ref{eq:D-K+result}).
The systematic error is the sum of the listed systematic errors added in quadrature.
Shifting the argument of the denominator slightly and following the same analysis steps, we obtain
Eq.~(\ref{eq:D-pi+result}).


\section{Conclusion and Discussion}
\label{sec:con}

To conclude, we provide the first lattice-QCD calculation of the form-factor ratio
$f^{(s)}_0(M^2_\pi)/f^{(d)}_0(M^2_K)$.
Our result leads to the factor $\mathcal{N}_F=1.094(88)(30)$, which is significantly closer to unity than the
sum-rule estimate \cite{Blasi:1993fi}, $\mathcal{N}_F^{\text{SR}}=1.24(8)$ (or
$\mathcal{N}_F^{\text{SR}}=1.3(1)$ \cite{Fleischer:2010ay}) used in previous analyses of hadronic
$f_s/f_d$~\cite{Fleischer:2010ca,2011arXiv1106.4435L}.
As noted above, the lack of significant $U$-spin breaking observed in this calculation is in accord with
other lattice-QCD calculations of similar form factors~\cite{Koponen:2011ev}.

We now examine how our new value of $\mathcal{N}_F$ affects the fragmentation-fraction ratio $f_s/f_d$.
LHC$b$ measures $f_s/f_d$ via $\mathrm{BR}(\bar{B}^0_s\to D_s^+\pi^-)/\mathrm{BR}(\bar{B}^0\to D^+K^-)$,
using the sum-rule estimate $\mathcal{N}_F^{\text{SR}}$, and finds
$f_s/f_d=0.250(24)_{\text{stat}}(17)_{\text{syst}}(17)_{\text{theo}}$ \cite{2011arXiv1106.4435L}.
Since $\mathcal{N}_F$ is not correlated with any other quantity in Eq.~(\ref{eq:fs_fd}), we easily find that
the fragmentation ratio should become 
\be
    \frac{f_s}{f_d} = 0.283(27)_{\text{stat}}(19)_{\text{syst}}(24)_{\text{theo}},
    \label{eq:hadronic1}
\ee
where the errors have also been scaled accordingly.
Superficially, our theoretical error is slightly larger than that obtained with the sum-rule 
estimate---8.5\% vs.~6.5\%.
Our error, however, is straightforward to improve, since it is dominated by Monte Carlo statistics, 
propagated through the chiral-continuum and $z$ extrapolations, as seen in Table~\ref{tab:f0_all}. 

Fleischer, Serra, and Tuning have proposed a second hadronic approach based on the ratio
$\mathrm{BR}(\bar{B}^0_s\to D_s^+\pi^-)/\mathrm{BR}(\bar{B}^0\to D^+\pi^-)$~\cite{Fleischer:2010ca}.
A complication is that a $W$-exchange diagram also contributes to the $\bar{B}_d^0\to D^+\pi^-$ decay,
leading to an additional factor $\mathcal{N}_E$ in the analog of Eq.~(\ref{eq:fs_fd}). 
It is estimated to be $\mathcal{N}_E=0.966(75)$~\cite{2011arXiv1106.4435L}. 
This method requires a similar input of the form-factor ratio
$\mathcal{N}'_F=[f^{(s)}_0)(M_\pi^2)/f^{(d)}_0(M_\pi^2)]^2$.
With our calculation, we can easily extrapolate the argument of the denominator, finding the form factor
ratio given in Eq.~(\ref{eq:D-pi+result}).
As a result, $\mathcal{N}'_F = 1.111(94)(34)$.
Reference~\cite{2011arXiv1106.4435L} uses the same sum-rule value $\mathcal{N}_F^{\text{SR}}=1.24(8)$ when
doing the analysis with similar approach, finding the fragmentation-fraction ratio to be 
$f_s/f_d=0.256(14)(19)(26)$.
We find that
\be
    \frac{f_s}{f_d} = 0.286(16)_{\text{stat}}(21)_{\text{syst}}(26)_{\text{latt}}(22)_{\text{NE}},
    \label{eq:hadronic2}
\ee
where the last two errors (major sources of the theoretical error) are shown explicitly.
The last error stems from the uncertainty in~$\mathcal{N}_E$.
The result Eq.~(\ref{eq:hadronic2}) agrees with that of the $D^+_s\pi^-/D^+K^-$ hadronic method,
Eq.~(\ref{eq:hadronic1}), and both agree with LHC$b$'s determination via a method employing semileptonic
decays, $f_s/f_d=0.268(8)_{\text{stat}}({}^{+24}_{-22})_{\text{syst}}$ \cite{Aaij:2011jp}, as well as the 
Particle Data Group's average of LEP and CDF, $f_s/f_d=0.288(24)$ \cite{Nakamura:2010zzi}.

As a by-product of the calculation, the form-factor ratio in Eq.~(\ref{eq:D-pi+result}) can be combined with 
factorization to estimate the ratio of branching ratios,
\begin{equation}
    \frac{\mathrm{BR}(\bar{B}^0_s\to D_s^+\pi^-)}{\mathrm{BR}(\bar{B}^0\to D^+K^-)} = 14.4\pm1.3,
\end{equation}
independently of experimental inputs except for quantities like $|V_{us}|f_K/|V_{ud}|f_\pi$ and lifetimes.
This ratio is consistent with the measured value $16\pm5$~\cite{Nakamura:2010zzi}, assuming no correlation
between the two processes.
Smaller experimental error bars would provide a better test of the validity of our calculation.

This work is based on only 4 out of 21 available MILC asqtad ensembles of lattice gauge configurations.
Further running on ensembles closer to the chiral and continuum limits will reduce the length of the 
extrapolations and, hence, control the growth through extrapolation of the statistical error.
At the current stage, however, the largest error in Eq.~(\ref{eq:hadronic1}) 
remains experimental statistics, stemming from the difficulty in reconstructing $D^{\pm}_s\to KK\pi$.

\acknowledgments

D.D. thanks Peter Lepage for his least square fitting codes, upon which some parts of his code are based. 
Computations for this work were carried out with resources provided by 
the USQCD Collaboration, the Argonne Leadership Computing Facility, 
the National Energy Research Scientific Computing Center, and the Los Alamos National Laboratory, which
are funded by the Office of Science of the United States Department of Energy; and with resources
provided by the National Institute for Computational Science, the Pittsburgh Supercomputer Center, 
the San Diego Supercomputer Center, and the Texas Advanced Computing Center, 
which are funded through the National Science Foundation's Teragrid/XSEDE Program. 
This work was supported in part by the U.S. Department of Energy under Grants 
No.~DE-FG02-91ER40628 (C.B.), 
No.~DOE~FG02-91ER40664 (D.D., Y.M.),
No.~DE-FC02-06ER41446 (C.D., J.F., L.L., M.B.O.), 
No.~DE-FG02-91ER40661 (S.G., R.Z.), 
No.~DE-FG02-91ER40677 (C.M.B, D.D, E.D.F., A.X.K.), 
No.~DE-FG02-04ER-41298 (J.K., D.T.); 
by the National Science Foundation under Grants 
No.~PHY-0555243, No.~PHY-0757333, No.~PHY-0703296 (C.D., J.F., L.L., M.B.O.), 
No.~PHY-0757035 (R.S.); 
by the URA Visiting Scholars' program (C.M.B., D.D., M.B.O.);
by the Fermilab Fellowship in Theoretical Physics (C.M.B.);
by the Science and Technology Facilities Council and the Scottish Universities Physics Alliance (J.L.);
by the MICINN (Spain) under grant FPA2010-16696 and Ram\'on y Cajal program (E.G.);
by the Junta de Andaluc\'ia (Spain) under grants FQM-101, FQM-330, and FQM-6552 (E.G.);
and by the Creative Research Initiatives program (3348-20090015) of the NRF grant funded by the Korean 
government (MEST) (J.A.B.).
This manuscript has been co-authored by employees of Brookhaven Science Associates, LLC,
under Contract No.~DE-AC02-98CH10886 with the U.S. Department of Energy. 
Fermilab is operated by Fermi Research Alliance, LLC, under Contract No.~DE-AC02-07CH11359 with
the U.S. Department of Energy.

\appendix*

\section{Staggered chiral perturbation theory for \boldmath $B\to D\ell\nu$ at nonzero recoil}
\label{app:rsChPT}

The material given in this Appendix extends the continuum-QCD $\chi$PT for 
$B\to D\ell\nu$~\cite{Chow:1993hr} to staggered fermions.
The staggered theory has 16 light pseudoscalar mesons for each meson of continuum QCD.
Their degeneracy is broken at finite lattice spacing with masses given at the leading order by
\cite{Bazavov:2009bb}
\be
    M^2_{qq'_\Xi} = \mu_0(m_q + m_{q'}) + a^2\Delta_{\Xi},
\ee
where $q,q'$ are the staggered quarks and $\mu_0$ is a continuum low-energy constant.
$a^2\Delta_{\Xi}$ are the splittings of the 16 mesons in lattice units, cf.\ Table~\ref{tab:chiralparameters}.
At this order in an expansion in $a^2$ they come in 5 multiplets, labeled $P$, $A$, $T$, $V$ and $I$, with
degeneracies $1$, $4$, $6$, $4$ and $1$, respectively.

In full (2+1) QCD rooted staggered chiral perturbation theory, we have the expression for~$h_+^{B\to D}$:
\begin{eqnarray}
    h_+^{\rm NLO}(w) = 1 & + & \frac{X_+}{m_c^2} + 
        \frac{\gDDp^2}{16\pi^2 f^2} \left[ \frac{1}{16}\sum_{\Xi} (2F^+_{\pi_\Xi} + F^+_{K_\Xi})
        - \frac{1}{2} F^+_{\pi_I} + \frac{1}{6}F^+_{\eta_I}  \right. \nonumber \\
    & + & a^2\delta'_V \left(
        \frac{M^2_{\pi_V}-M^2_{S_V}}{(M^2_{\pi_V} - M^2_{\eta_V})(M^2_{\pi_V} - M^2_{\eta'_V})} F^+_{\pi_V} 
      + \frac{M^2_{\eta_V}-M^2_{S_V}}{(M^2_{\eta_V} - M^2_{\eta'_V})(M^2_{\eta_V} - M^2_{\pi_V})} F^+_{\eta_V}
    \right.  \nonumber \\
    & & + \left.\left. 
        \frac{M^2_{\eta'_V}-M^2_{S_V}}{(M^2_{\eta'_V} - M^2_{\eta_V})(M^2_{\eta'_V}- M^2_{\pi_V})} F^+_{\eta'_V}
        \right) + (V\rightarrow A) \right].
    \label{eq:xlog_bd}
\end{eqnarray}
Similarly for $h^{(s)}_+$ ($B_s \to D_s$), we have 
\begin{eqnarray}
    h_+^{(s),\rm NLO}(w) = 1 & + & \frac{X_+}{m_c^2} + 
        \frac{\gDDp^2}{16\pi^2 f^2} \left [ \frac{1}{16}\sum_{\Xi} (F^+_{S_\Xi} + 2F^+_{K_\Xi})  -  
            F^+_{S_I} + \frac{2}{3}F^+_{\eta_I}  \right. \nonumber \\
    & + & a^2\delta'_V \left(\frac{M^2_{S_V}-M^2_{\pi_V}}{(M^2_{S_V}- M^2_{\eta_V})(M^2_{S_V} - M^2_{\eta'_V})} F^+_{S_V} +
        \frac{M^2_{\eta_V}-M^2_{S_V}}{(M^2_{\eta_V} - M^2_{\eta'_V})(M^2_{\eta_V} - M^2_{\pi_V})} F^+_{\eta_V}  
        \right.  \nonumber \\
    && + \left.\left.
        \frac{M^2_{\eta'_V}-M^2_{\pi_V}}{(M^2_{\eta'_V} - M^2_{\eta_V})(M^2_{\eta'_V}- M^2_{S_V})} F^+_{\eta'_V}
        \right)  + (V\rightarrow A) \right],
    \label{eq:xlog_bs}
\end{eqnarray}
where the masses of the flavor-taste singlet mesons $\eta_I$ and nonsinglet mesons 
$\eta_{V(A)}$, $\eta'_{V(A)}$ are given by~\cite{Laiho:2005ue}
\ben
M_{\eta_I}^2 &=& \frac{1}{3}\left( M_{\pi_I}^2 + 2 M_{S_I}^2 \right), \nonumber\\
M_{\eta_{V}}^2 &=& \frac{1}{2}\left( M_{\pi_{V}}^2 + M_{S_{V}}^2 + \frac{3}{4} a^2\delta'_{V} - Z\right), \nonumber  \\
M_{\eta'_{V}}^2 &=& \frac{1}{2}\left( M_{\pi_{V}}^2 + M_{S_{V}}^2 + \frac{3}{4} a^2\delta'_{V} + Z\right), \nonumber \\
Z &=& \left[ (M_{S_{V}}^2-M_{\pi_{V}}^2)^2 - \frac{1}{2}a^2 \delta'_{V}(M_{S_{V}}^2-M_{\pi_{V}}^2) + \frac{9}{16}(a^2 \delta'_{V})^2 \right]^{1/2}, \nonumber \\
&& (V\rightarrow A).
 \label{eq:eta}
\een
In Eqs.~(\ref{eq:xlog_bd}) and~(\ref{eq:xlog_bs}), $F^+_j$ is short for the function 
$F^+(w, M_j,\Delta^{(c)}/M_j)$, defined by
\be
    F^+(w,m,x) = -2\left[ (w+2) I_1(w,m,x) + (w^2-1)I_2(w,m,x) -\frac{3}{2}I_3(w,m,x) - 
        \frac{3}{2}I_3(w,m,0) \right],  \label{eq:F+}
\ee
where 
\be
    I_i(w,M,x) = - \left[ M^2 x E_i(w) + M^2 x^2 \ln \left(\frac{M^2}{\Lambda^2}\right ) G_i(w) + 
        M^2x^2 F_i(w,x) \right]
\ee
and the functions $E, G$ are given by
\begin{subequations}
\begin{eqnarray}
    E_1(w) &=& \frac{\pi}{w+1},  \\
    E_2(w) &=& \frac{-\pi}{(w+1)^2},  \\
    E_3(w) &=& \pi ,  \\
    G_1(w) &=& \frac{-1}{2(w^2-1)} [w - r(w) ],  \\
    G_2(w) &=& \frac{1}{2(w^2-1)^2} [w^2+2 -3w\,r(w) ],  \\
    G_3(w) &=& -1,
\end{eqnarray}    
\end{subequations}
with  
\be
    r(w) = \frac{1}{\sqrt{w^2 -1} }\ln (w + \sqrt{w^2 -1} ).
\ee
The functions $F_i$ are given by
\begin{subequations}
\label{appendix:functions1}
\begin{eqnarray}
    F_1(w,x) &=& \frac{1}{x^2}\int_0^{\pi/2}d\theta \frac{a}{1+w \sin2\theta} \left\{
        \pi\left(\sqrt{1-a^2}-1\right) - 2\left[f(a) -a \right] \right \},  \\
    F_2(w,x) &=& \frac{1}{x^2}\int_0^{\pi/2}d\theta \frac{a \sin 2\theta}{(1+w \sin2\theta)^2} \left\{
        -\frac{3\pi}{2}(\sqrt{1-a^2} -1) + \frac{\pi a^2}{2\sqrt{1-a^2}} 
        \right. + \nonumber \\ & & \left. 
        \frac{3-4a^2}{1-a^2} f(a) -3a \right \}, \\
    F_3(w,x) &=& \frac{1}{x} \left\{ \pi\left(\sqrt{1-x^2} -1\right) - 2[f(x)-x]\right \}, 
\end{eqnarray}    
\end{subequations}
where 
\be
a = \frac{x \cos\theta}{\sqrt{1+ w\sin 2\theta}},
\ee
\be
    f(x) = \left\{
    \begin{array}{ll}
        \sqrt{1-x^2}\tan^{-1}\left[x/\sqrt{1-x^2}\right],            & |x|<1,\\
        \frac{1}{2}\sqrt{x^2-1}\ln\left[1-2x(x+\sqrt{x^2-1})\right], & |x|>1,
    \end{array}
    \right.
\ee
and $x = \Delta^{(c)}/M_j$ where $M_j$ is the corresponding meson mass.
The $D_{(s)}$-$D_{(s)}^*$ splittings are $\Delta^{(c)}=140.6$~MeV, $\Delta^{(c)}_s=143.9$~MeV.
When the continuum and chiral limits are taken, the taste splittings vanish, and the 16 lowest $M_j$s all
tend to the physical pion mass, which is around $135$~MeV.
The extrapolation to the physical pion mass switches from $|x|<1$ to $|x| > 1$, requiring both expressions
for $f(x)$.

\bibliographystyle{apsrev4-1}
\bibliography{Bs_project}

\end{document}